\documentclass[twocolumn]{aastex631}

\newcommand{\teq}{$T_{\rm eq}$}
\newcommand{\rev}{}

\usepackage{amsmath}
\usepackage{CJK}
\shorttitle{The Atmospheric Circulation of Cloudy Hot-Jupiters}
\shortauthors{Tan et al.}

\begin{document}
\begin{CJK*}{UTF8}{gbsn}

\title{The Atmospheric Circulation of Cloudy Hot and Ultra-hot Jupiters: the Need for Dissipation of Large-scale Flows to Reduce the Nightside Brightness Temperatures of Ultra-hot Jupiters}

\author[0000-0003-2278-6932]{Xianyu Tan (谭先瑜)}
\affiliation{Tsung-Dao Lee Institute, Shanghai Jiao Tong University, 1 Lisuo Road, Shanghai 200127, China}
\affiliation{School of Physics and Astronomy, Shanghai Jiao Tong University, 800 Dongchuan Road, Shanghai 200240, China}
\affiliation{State Key Laboratory of Dark Matter Physics, Shanghai Jiao Tong University, 1 Lisuo Road, Shanghai 200127, China}
\email{xianyut@sjtu.edu.cn}

\author[0000-0001-9700-9121]{Yanhong Lai (赖燕红)}
\affiliation{Tsung-Dao Lee Institute, Shanghai Jiao Tong University, 1 Lisuo Road, Shanghai 200127, China}

\author[0009-0002-1101-2798]{Jinghao Zhang (张景灏)}
\affiliation{Department of Astronomy, School of Physics, Peking University, Beijing 100871, China}
\affiliation{Kavli Institute for Astronomy and Astrophysics, Peking University, Beijing 100871, China}

\author[0000-0002-9258-5311]{Thaddeus D. Komacek}
\affiliation{Atmospheric, Oceanic, and Planetary Physics, Department of Physics; University of Oxford, Oxford OX1 3PU, UK} 

\author[0000-0003-3431-8570]{Tao Cai (蔡涛)}
\affiliation{State Key Laboratory of Lunar and Planetary Sciences, Macau University of Science and Technology, Macau, China} 
\affiliation{Zhuhai MUST Science and Technology Research Institute, Zhuhai, China}



\begin{abstract}
We perform general circulation modeling of hot and ultra-hot Jupiters across equilibrium temperatures of $T_{\rm eq}=$1600 to 2800 K, incorporating cloud formation and its feedback on atmospheric dynamics. This study specifically examines whether clouds explain the systematically low nightside brightness observed in Spitzer thermal phase curves of ultra-hot Jupiters. We found that clouds generally increase phase-curve amplitudes and reduce eastward offsets, and these effects are most pronounced at relatively low $T_{\rm eq}$ and gradually weaken as $T_{\rm eq}$ increases. We discussed two mechanisms responsible for reducing the phase-curve offsets by clouds: the dynamical and geometric effects.  Without explicit dissipation of large-scale winds, strong equatorial jets develop, maintaining a hot equatorial region and clearing clouds near the nightside equator at $T_{\rm eq}\gtrsim 2200$ K, with thick clouds instead forming mainly at mid-to-high latitudes. These patchy nightside clouds do not sustain low nightside fluxes as temperatures rise, suggesting that clouds alone cannot explain the observed nightside brightness for ultra-hot Jupiters.  {\rev Strong dissipation, parameterized as magnetic drags, suppresses equatorial jets and promotes nightside cloud coverage. In both clear and cloudy models, drag weakens the dependence of Spitzer nightside brightness temperature on equilibrium temperature. At a given equilibrium temperature, clouds systematically lower the nightside brightness temperature relative to cloud-free models with the same drag prescription.} Clouds also dampen amplitudes of thermal spectra on the nightside and, under small-particle conditions, can even produce emission features. Due to limited dayside cloud cover, our models yield negligible Bond albedos. Finally, interpreting simulated thermal phase curves with energy balance models {\rev in near-IR and IR bands} tends to overestimate the Bond albedos of ultra-hot Jupiters by an order of magnitude or more.
\end{abstract}

\keywords{ Exoplanet atmospheres (487); Exoplanet atmospheric
dynamics (2307); Exoplanet atmospheric structure (2310);
Planetary atmospheres (1244); Hot Jupiters (753)}


\section{Introduction} \label{sec:intro}

Hot Jupiters, a class of giant exoplanets in close orbits around their host stars, are among the most intensively studied exoplanetary groups due to their observational accessibility and peculiar characteristics. Atmospheric characterization offers critical insights into the formation and evolutionary pathways of close-in exoplanets as a population, as well as into exotic atmospheric processes not expected in their colder counterparts \citep{Madhusudhan2019,fortney2021}.

Phase-curve observations---which monitor the system’s flux throughout the planet's orbit---constitute one of the most comprehensive methods for characterizing the multidimensional nature of planetary atmospheres \citep{Parmentier2018review}. Thermal phase curves trace the zonal (east–west) distribution of brightness temperatures, while short-wavelength phase curves reveal the longitudinal patterns of reflective materials, such as high-temperature condensates or photochemical hazes. Spectral measurements at multiple orbital phases, enabled by facilities like HST and JWST and future missions such as Ariel, further constrain the global thermal structure and chemical processes. 

Across these observations, clouds are of critical importance. Their high opacity and strong radiative heating/cooling effects substantially influence atmospheric dynamics, thermal structure, chemical transport, and spectroscopic signatures \citep{showman2020}. {\rev One observational motivation for investigating nightside clouds comes from mid-infrared phase curves obtained with the Spitzer Space Telescope.} These data reveal that the nightside brightness temperatures of hot Jupiters at 3.6- and 4.5-$\mu$m Spitzer bands cluster around $\sim1100$ K, showing a somewhat flat trend across a wide range of planetary equilibrium temperatures \citep{beatty2019,keating2019,bell2021,dang2025}. {\rev Even in cloud-free models, the competition between heat transport and radiative cooling may produce a weak dependence of nightside temperature on equilibrium temperature \citep[e.g.,][]{komacek2018rnaas,parmentier2021}. Persistent, optically thick nightside clouds can lower the brightness temperature by shifting the emitting level to lower pressures, with little change in the slope of this trend \citep{parmentier2021}. Although a nearly constant cloud-top temperature has been suggested as an explanation for the observed clustering \citep{keating2019}, it is unclear what would maintain such a temperature as irradiation varies. If nightside clouds become patchier at higher equilibrium temperatures \citep[e.g.,][]{komacek2022}, emission from deeper, hotter regions can steepen the trend. This raises the question of whether clouds forming in a three-dimensional circulation can reproduce both the low Spitzer nightside brightness temperatures and their weak dependence on equilibrium temperature.}

A broad range of modeling work has employed three-dimensional (3D) general circulation models (GCMs) to investigate the influence of clouds on the atmospheric dynamics of hot Jupiters and their observational signatures. These studies range from parametric surveys exploring broad trends \citep[e.g.,][]{roman2021,parmentier2021, Helling2023, kennedy2025} to simulations aimed at explaining specific observations of individual planets \citep[e.g.,][]{lee2016,lines2018,lee2023hj, teinturier2024,noti2024,kiefer2024,bell2024wasp43b,Mehta2026,Mehta2026b}. Other efforts have incorporated simplified dynamical transport and inhomogeneous structures into detailed microphysical calculations \citep[e.g.,][]{powell2018,powell2024}.

A general qualitative conclusion from these studies is that clouds tend to reduce the nightside brightness temperature. However, the observed low and nearly constant nightside temperatures in the Spitzer mid-IR bands deserve a more quantitative investigation using GCMs. Previous GCM studies that are related to this issue have often relied on one or more simplifying assumptions, such as: static clouds that are permitted only by thermodynamic conditions but do not interact with atmospheric flows; radiatively passive clouds whose opacity are excluded from radiative transfer calculations in the GCM; simplified radiative transfer schemes that inaccurately compute heating/cooling rates and thermal structures; and certain inconsistencies between the GCM and radiative transfer post-processing. These limitations have precluded a fully rigorous, population-level comparison between cloudy GCMs and the Spitzer phase-curve observations.

{\rev An additional process important in ultra-hot Jupiters is the dissociation and recombination of molecular hydrogen. Dissociation absorbs energy on the hot dayside, which is transported by atomic hydrogen and released through recombination in cooler regions, enhancing day-to-night energy transport and reducing the thermal contrast \citep{bell2018,komacek2018rnaas,tan2019uhj,mansfield2020,roth2021,tan2024}. Because this modifies the temperatures that govern cloud formation, understanding nightside emission in ultra-hot Jupiters requires considering this process together with cloud radiative feedback and large-scale flow dissipation. These have been partially considered in semi-grey GCMs for individual planets \citep{komacek2022}, but not yet in systematic parametric modeling.}

The first goal of this work is to investigate the influence of clouds on atmospheric circulation and to test cloud effects more rigorously against the observed Spitzer nightside brightness temperatures across a broad range of equilibrium temperatures for hot and ultra-hot Jupiters (UHJs). We employ a GCM with a self-consistent cloud formation scheme, a realistic radiative transfer module, thermodynamics of hydrogen dissociation and recombination, and consistent post-processing radiative transfer. Our cloud scheme accounts for formation and evaporation under thermodynamic equilibrium, gravitational settling, transport by large-scale flows, and the radiative feedback of clouds to the atmospheric radiative transfer. This approach allows the resulting cloud distributions and atmospheric structures to fully capture cloud-circulation interactions. The opacity and chemistry setups of the radiative transfer post-processing are consistent with the GCM, ensuring an accurate representation of brightness temperatures from the simulated cloud dynamics. Similar modeling frameworks have recently advanced the understanding of variability and near-IR colors in substellar atmospheres \citep{tan2025,Teinturier2026,Mehta2026}; here, we extend their application to hot Jupiter atmospheres.

As discussed in Section \ref{sec:results}, we find that our cloudy GCMs without large-scale flow dissipation predict nightside brightness temperatures significantly higher than those observed in the Spitzer 4.5-$\mu$m band at equilibrium temperatures $\geq2200$ K. Consequently, the second goal of this work is to explore how dissipation---primarily magnetohydrodynamic (MHD) dissipation---can further reduce the nightside temperature. As atmospheric temperature rises, increased conductivity due to metal ionization enhances interactions between the bulk flow and the planetary magnetic field. These interactions are expected to limit the kinetic energy of large-scale flows \citep{Batygin2010,perna2010,Menou2012} and reshape the radiative-forcing-driven jets and waves, potentially altering heat transport, phase-curve properties, cloud formation, and disequilibrium chemistry \citep[e.g.,][]{beltz2022}. Previous studies in this area have either simulated full MHD while simplifying other atmospheric physics \citep[e.g.,][]{rogers2014,rogers2017,boning2025} or incorporated atmospheric physics while parameterizing magnetic interactions as a drag in hydrodynamic models \citep[e.g.,][]{rauscher2013,arcangeli2019}. In this study, we parameterize MHD dissipation within our GCM as drag and present the results in Section \ref{sec:drag}.

The paper is organized as follows. Section \ref{sec:model} introduces the GCM, post-processing tools, and simulation setups. Section \ref{sec:results} presents results from models without drag. Section \ref{sec:drag} describes models with additional drag and compares them to the drag-free cases. Section \ref{sec:dis} discusses the implications for nightside spectra and Bond albedo, as well as values derived from energy balance models. Finally, Section \ref{sec:con} summarizes the key findings.

\section{Model Description} \label{sec:model}
\subsection{The general circulation model: \texttt{ADAM}}
This work uses ADvanced Atmospheric MITgcm (\texttt{ADAM}) with SPARC, H$_2$-dissociation, and active tracer clouds, to simulate the global atmospheric circulation, thermal structure and cloud distributions in a dynamically and radiatively self-consistent manner. \texttt{ADAM} refers to an umbrella designation for exoplanet modeling frameworks built upon the MITgcm \citep{adcroft2004} and the non-grey radiative transfer code of \cite{marley1999}. Under this naming convention, model configurations are referenced by specifying the relevant physical modules. We clarify that \texttt{ADAM} refers specifically to the lineage of models descending from the developments initiated by \cite{showman2009} and extended through a series of key advancements afterwards \citep[e.g.,][]{lewis2010,kataria2013,parmentier2013,komacek2017,tan2019uhj,tan2021bd1,tan2021bd2,komacek2022,steinrueck2023,Mehta2026}. More details are referred to \cite{Mehta2026}. 

The dynamical core solves the standard primitive equations of dynamical meteorology, including equations of the horizontal angular momentum conservation, hydrostatic balance in the radial direction, mass continuity, thermodynamics and ideal-gas equation of state for hydrogen-helium dominated atmospheres. The equations are solved using the cubed-sphere grid \citep{adcroft2004}. Radiative heating and cooling by stellar irradiation and thermal fluxes of the planet are the main drivers of the large-scale circulation. We adopt the radiative transfer model originally developed in \cite{marley1999} and coupled to the MITgcm in \cite{showman2009}, which utilizes the equilibrium gas chemistry correlated-k method to compute the wavelength-dependent (non-grey) radiative transfer. {\rev The correlated-k opacity tables include TiO and VO, and these strong optical absorbers can drive dayside thermal inversions \citep[e.g.,][]{hubeny2003,fortney2008unified,roth2024}, alongside other absorbers important in ultra-hot Jupiter atmospheres \citep{lothringer2018}. The resulting heating may substantially suppress dayside cloud formation in our modeled temperature range. Retaining TiO and VO allows us to explore this regime of strong dayside absorption, motivated by observations of dayside thermal inversions among ultra-hot Jupiters \citep[e.g,][]{Haynes2015,Mikal-Evans2020, mansfield2021,Coulombe2023}. More details of the radiative-transfer setup for this work are referred to \cite{tan2024}. }

\subsection{The hydrological cycles}
The ``hydrological cycle" broadly refers to tracer-based processes that are materially advected by flows and with sources and sinks such as conversions between different tracers and gravitational sedimentation of cloud particles. The hydrological cycles may include general chemical processes, and here we include the cloud hydrological cycle and hydrogen dissociation and recombination, as described below. 

In the first hydrological cycle of cloud formation, we include two relatively high-temperature condensing species, MgSiO$_3$ and Al$_2$O$_3$ which are among the most abundant condensates in terms of mass mixing ratio at relevant temperatures of hot Jupiters \citep{visscher2010,wakeford2017,gao2021}. Including more condensing species moderately increases the total cloud opacity which is expected to yield quantitatively different results but should not qualitatively influence our conclusions. For each condensing species, one tracer represents the mass mixing ratio of cloud particles, $q_c$, and another represents the ``vapor'' form, $q_v$. Unlike water clouds in Earth's atmosphere, which form from the phase change of a single constituent, mineral cloud formation involves complex gas- and solid-phase chemical reactions \citep{visscher2006,helling2019}. In principle, there is no single constituent as the appropriate ``vapor''. Nevertheless, our model considers feedback from only cloud particles but vapor is passive; one vapor tracer is considered for one condensing species for dynamical consideration (because the spatial distribution of the gas is also crucial for cloud formation) and also for the simplicity of the model. Latent heat and effects of change of mean molecular weight associated with condensation/evaporation are neglected as they are much smaller than the radiative feedback at relevant conditions \citep{tan2017,tan2019}. We assume homogeneous cloud formation \citep{ackerman2001} using condensation curves shown in \cite{visscher2006,visscher2010} and without considering mix-material grain growth \citep{helling2019,lee2023hj}. Similar cloud-hydrological schemes have been used in our recent work \citep{tan2021bd2,komacek2022,tan2025,Mehta2026,lai2026,lai2026bd}.

Once condensation occurs, the vapor tracer turns into the cloud tracer over a characteristic timescale $\tau_c\sim 100$ s, and the cloud particles are assumed to be in a log-normal distribution to mimic the broad distribution driven by condensational and collisional growth \citep{gao2018,gao2021,powell2024}:
\begin{equation}
    n(r) = \frac{\mathcal{N}_c}{\sqrt{2\pi}\sigma r}\exp \left(-\frac{[\ln (r/r_0)]^2}{2\sigma^2}\right),
\end{equation}
where $\mathcal{N}_c$ is the total cloud particle number per unit mass, $n(r) = d\mathcal{N}_c/dr$ is the number density distribution, $\sigma$ is a nondimensional width and $r_0$ is the reference particle radius. We use a fixed $\sigma=0.5$ to represent the potentially wide size distribution following \cite{ackerman2001,tan2019,lee2024bd}. The reference radius $r_0$ is a free parameter to control clouds' optical depth and settling flux. The cloud tracer mass mixing ratio is related to $\mathcal{N}_c$ via $q_c=\frac{4}{3}\pi \rho_c \mathcal{N}_c r_0^3\exp (9\sigma^2/2)$, where $\rho_c$ is the density of solid cloud material. 

Cloud particles decouple from the gas in the vertical direction under gravity and settle at a terminal speed as a function of temperature, pressure, and particle size. We calculate the terminal speed of particles based on the formula summarized in \cite{ackerman2001,parmentier2013}, at different $r$ with 40 bins between 0.1 to 100 $\mu$m and integrate over the size distribution with particle mass to obtain the settling flux, which is included as sinks and sources in the cloud tracers. The extinction optical depth, single scattering albedo, and scattering asymmetry factor of clouds as a function of wavelength based on the spatial and time-dependent cloud tracer are interpolated from grids of pre-existing Mie calculations of homogeneous, spherical particles \citep{ackerman2001}. For optical calculations, 55 radius bins are used between 0.001 to 300 $\mu$m over the cloud size distribution. 

The second hydrological cycle is the hydrogen dissociation and recombination in the UHJ regime \citep{bell2018,komacek2018rnaas}. Our tracer-based approach is based on that of \cite{tan2019uhj} with updates in \cite{komacek2022}. The mass mixing ratio of the atomic hydrogen H relative to the total gas of H+H$_2$+He is treated as a tracer. The sink and source of the tracer are included as H$_2$-H conversion, calculated from the Saha equation for the H$_2$-H pair. The chemical ``latent heat'' of the H$_2$-H reaction and the change of mean molecular weight is included via the thermodynamics, hydrostatic balance and ideal-gas equation of state in the primitive equation set. A modified potential temperature associated with the H$_2$-H conversion is implemented to include the above effects \citep{tan2019uhj}. A soft adjustment scheme with a finite H$_2$-H conversion timescale of 40 s is implemented to ensure numerical stability \citep{tan2024}.

\subsection{Radiative Transfer Post-Processing}
Offline radiative transfer calculations based on 3D GCM outputs of temperature and cloud fields with higher spectral resolutions are required for spectra and brightness temperature predictions. Our post-processing is based on \texttt{PICASO}, an open-source radiative transfer code for computing the reflected, thermal, and transmission spectrum of planets and brown dwarfs \citep{batalha2019, Mukherjee2023}. \texttt{PICASO} shares the same opacity source and chemistry routines as the SPARC/MITgcm and guarantees full consistency between the GCM and radiative transfer post-processing. This can rigorously test the effects of clouds and heat transport on the spectra and wavelength-dependent phase curves. The source function technique based on \cite{toon1989} in \texttt{PICASO} that reconstructs the intensity field based on the flux field naturally considers the limb darkening or brightening effects of planetary thermal emission.

\subsection{Model Grid Parameters and Setup}
We conduct grids of GCM simulations over a wide range of equilibrium temperature $T_{\rm eq}=1600$ to 2800 K with a step of 200 K. Here  $T_{\rm eq}$ is defined as a blackbody temperature required to radiate away stellar irradiation energy over the full globe, assuming zero Bond albedo. We assume a host star with a stellar effective temperature of 6500 K, a mass of 1.34 solar mass, and a stellar radius of 1.5 solar radii. This setup is identical to that in one of the model branches of \cite{tan2024}. {\rev For comparison, host-star effective temperatures in the Spitzer phase-curve sample of \cite{dang2025} span approximately 4100 to  10200 K, with a median near 6000 K; for those with planetary $T_{\rm eq}>2000$, the median stellar effective temperature is 6415 K. We adopt a single stellar type to examine systematic trends with irradiation, while recognizing that variations in stellar spectra and the corresponding synchronous rotation periods can contribute to differences between our model sequence and the observed population.} By assuming the synchronous rotation of close-in planets, a lower $T_{\rm eq}$ implies a larger orbital separation and thus a longer planetary rotation period. For $T_{\rm eq}=$1600, 1800, 2000, 2200, 2400, 2600 and 2800 K of the grid, the planetary rotation period (also the orbital period) is set to be about 4.37, 3.07, 2.24, 1.68, 1.30, 1.06 and 0.82 days, respectively. We assume an atmospheric metallicity of $1\times$solar and equilibrium chemistry throughout the grid. The planetary surface gravity of all cases is assumed to be 10 ${\rm ms^{-2}}$ and the planetary radius is assumed to be $10^8$ m, typical for inflated hot Jupiters. {\rev The planetary radius and surface gravity in the primitive equation sets and thermal emission post-processing are treated as constants, so they are not specified at any pressure levels.} We employ the cubed-sphere grid with a horizontal resolution of C32, equivalent to about $2.8^{\circ}\times2.8^{\circ}$ in longitude and latitude. In the vertical pressure coordinate, we use 53 layers between 100 to $5\times10^{-6}$ bars. 

\section{Drag-free GCMs} \label{sec:results}
\subsection{The General Circulation and Phase-Curve Amplitudes}

\begin{figure*}
    \begin{center}
    \includegraphics[width=0.9\textwidth]{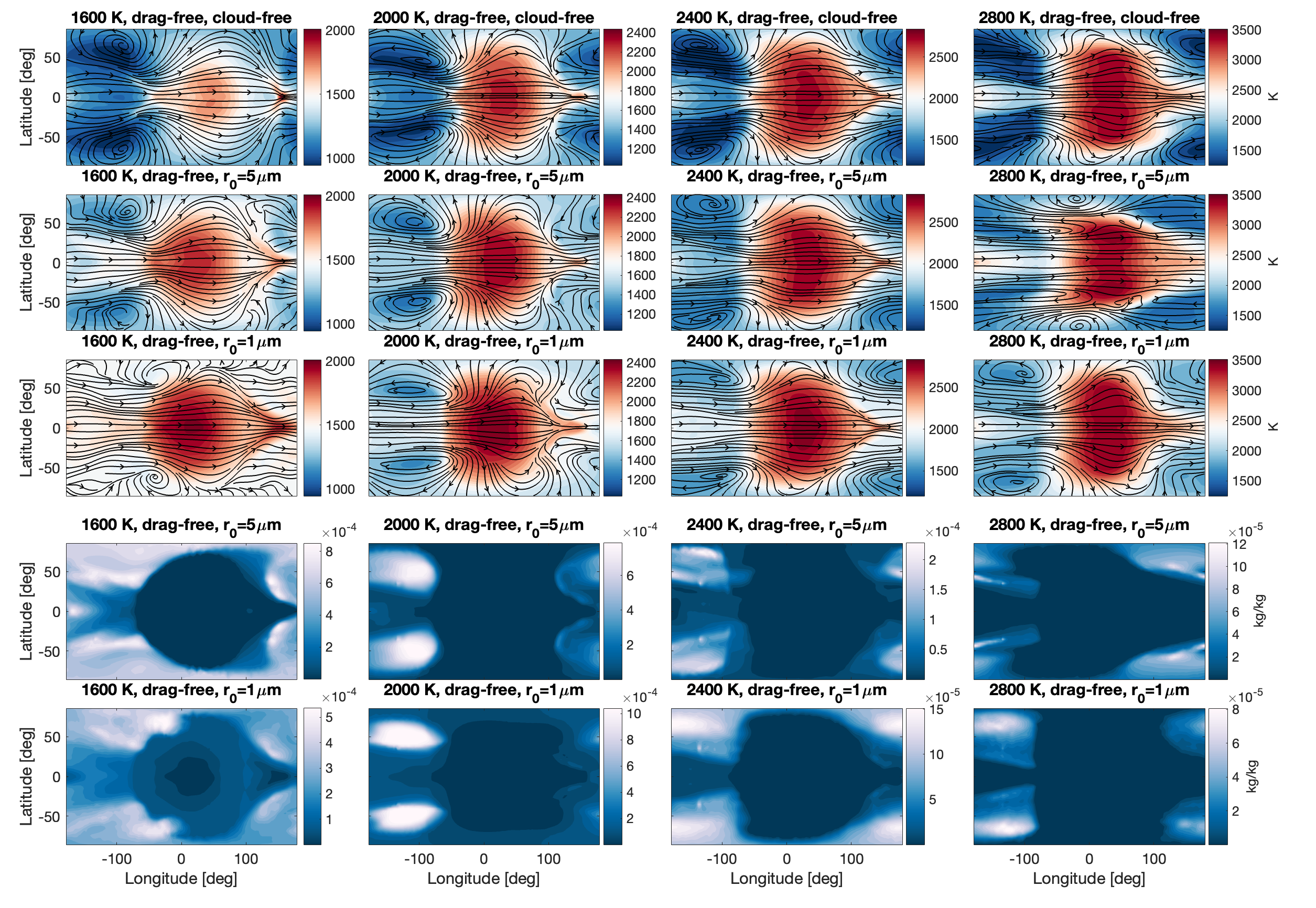}
    \caption{{\it Top three rows}: temperature maps at 32 mbar for a subset of drag-free models with varying planetary equilibrium temperature $T_{\rm eq}$ from 1600 K to 2800 K (from left to right). Horizontal winds are represented by streamlines. The top row shows cloud-free models; the second row shows cloudy models with a reference particle radius of $r_0=5~\mu$m; and the third row shows cloudy models with $r_0=1~\mu$m. {\it Bottom two rows}: vertically averaged total cloud mass mixing ratio (MgSiO$_3$ and Al$_2$O$_3$, counting from 0.1 bar to top of the model), in a unit of kg/kg, as a function of longitude and latitude for cloudy models.  These results are time-averaged over the last 200-day simulations.}
    \label{fig:maps}
    \end{center}
\end{figure*}

By comparing cloud-free to cloudy results, we may quantitatively access effects of clouds on the circulation. Figure \ref{fig:maps} shows the horizontal temperature maps at 32 mbar, a pressure level close to the thermal photosphere, for a subset of cloud-free, cloudy models with a reference radius $r_0=5~\mu$m and $r_0=1~\mu$m. To the first order, all models show qualitatively consistent features consisting of a prominent equatorial eastward jet and stationary global-scale waves which are shifted eastward relative to the sub-stellar longitude. These are the canonical circulation patterns that emerge from the strongly irradiated, moderately rotating, tidally locked hot Jupiter atmospheres \citep{heng2015,showman2020}. 
There are two main quantitative differences between cloud-free and cloudy models, however. As visually seen in the vertical direction of the first three rows of Figure \ref{fig:maps}, adding clouds results in, first, an obvious greenhouse warming at the same pressure level, where both dayside and nightside temperatures increase. The cloud warming effect is stronger in the case of $r_0=1~\mu$m than that of $r_0=5~\mu$m because of the higher cloud opacity in the former case. The cloud greenhouse warming, especially on the nightside, are effective at all \teq. Second, the degree of eastward shift of the thermal pattern relative to the substellar point decreases with inclusion of clouds, and this is more prominent at lower \teq$\lesssim 2400$ K. At a given \teq, cases with $r_0=1~\mu$m has less eastward shifts than those of $r_0=5~\mu$m.

The vertically-averaged cloud distributions for a subset of models are shown in the lower two rows in Figure \ref{fig:maps}. At all \teq, there are fewer clouds on the dayside due to the high temperatures. At relatively lower temperatures of the modeled range, clouds are well covered on the nightside. However, as temperatures increase, clouds become significantly more heterogeneous on the nightside, showing a depletion at the equator but significant coverage at mid-high latitudes where they can be thermodynamically maintained in cold regions associated with Rossby wave-induced gyres. The depletion of clouds at the equator on the nightside is likely due to the hot equator maintained by the strong equatorial jets (see the top three rows).

\begin{figure*}
    \begin{center}
    \includegraphics[width=0.9\textwidth]{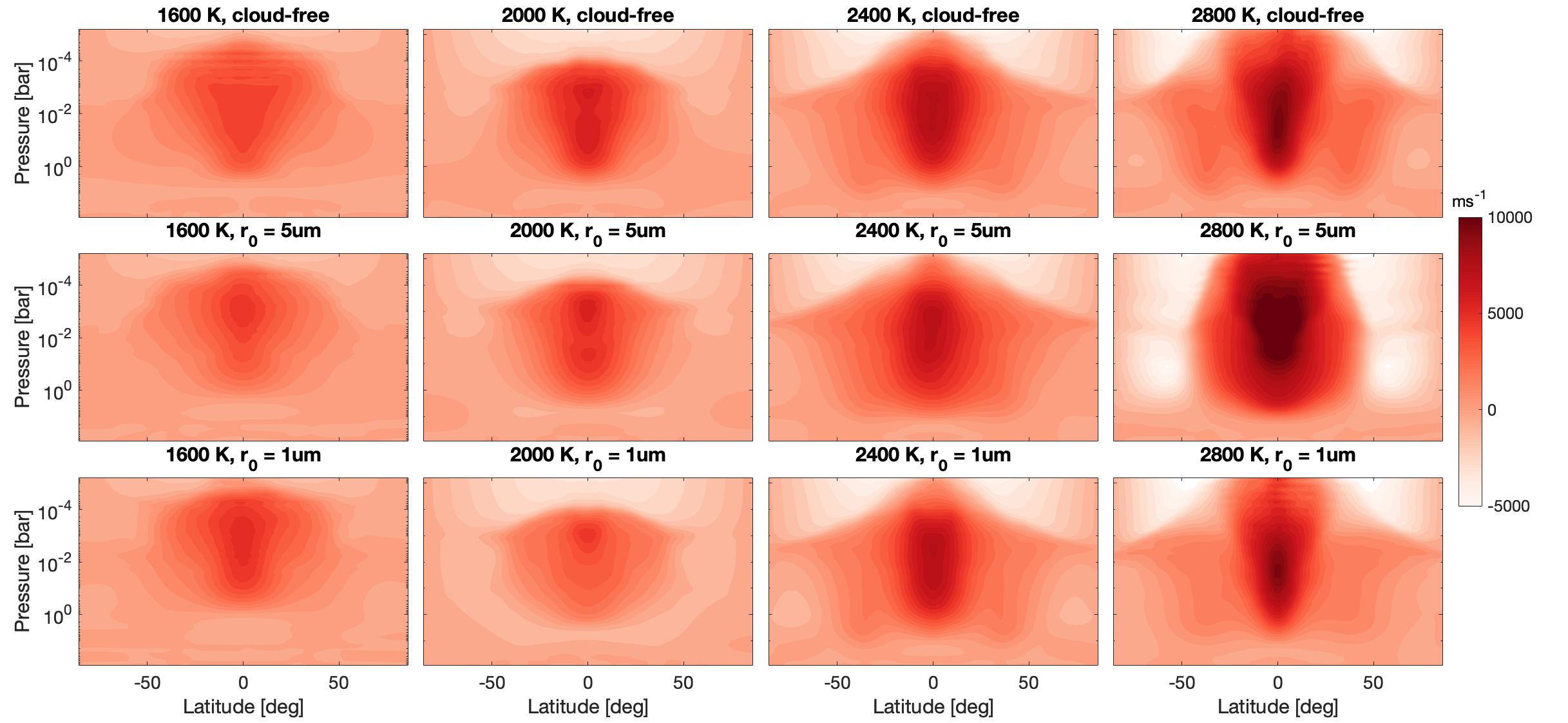}
    \caption{Zonal-mean zonal winds for drag-free models without clouds (first row), with clouds of $r_0=5~\mu$m (second row) and clouds of $r_0=5~\mu$m (third row). Results are time-averaged over the last 200-day simulations.}
    \label{fig:jets}
    \end{center}
\end{figure*}

All models show a dominant equatorial superrotating jet with speeds from a few km~s$^{-1}$ up to about 10 km~s$^{-1}$ and most of the jets extend from nearly the model top  down to much deeper than typical thermal photospheres, similar to results in \cite{tan2024}. Figure \ref{fig:jets} shows the zonal-mean zonal wind as a function of latitude and pressure from a collection of our model grids of drag-free but without clouds and with clouds of different particle sizes.  At a given equilibrium temperature, the zonal jets remain qualitatively similar among models without and with clouds of different sizes. The jet speed increases with the equilibrium temperature $T_{\rm eq}$ but shows small variation among different cases at a given $T_{\rm eq}$.  Only subtle differences are observed in some cases. For example, at $T_{\rm eq}=2800$ K, the high-latitude westward mean flow, merely a reflection of the high-latitude Rossby wave patterns in a zonal-mean sense, appears to be stronger and more coherent in the vertical direction in the case with clouds of $r_0=5~\mu$m than in the other two cases.

\begin{figure*}
    \begin{center}
    \includegraphics[width=1.\textwidth]{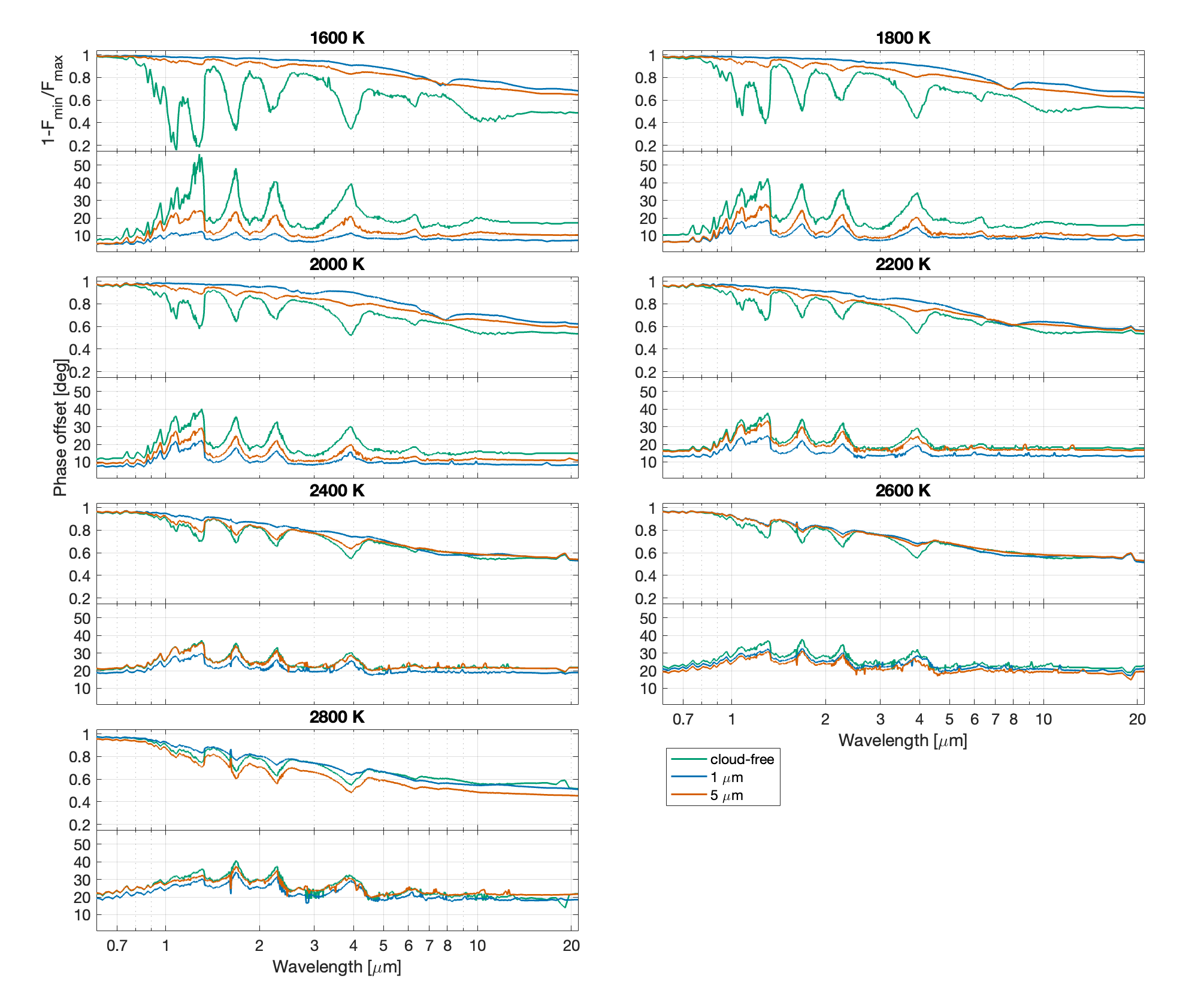}
    \caption{Thermal phase-curve amplitudes $1-F_{\rm min}/F_{\rm max}$ (upper part of each panel) and eastward offsets (bottom part of each panel) as a function of wavelength for drag-free models with different \teq. {\rev Green denotes the cloud-free sequence, vermilion the $5~\mu$m cloud sequence, and blue the $1~\mu$m cloud sequence.}}
    \label{fig:offset}
    \end{center}
\end{figure*}

Changes in circulation and the inclusion of clouds substantially influence the amplitudes and offsets of thermal phase curves.   We calculated thermal phase curves based on outputs of cloud-free and cloudy GCMs on the temperature grid, and the phase-curve flux amplitude and offsets as a function of wavelength are shown in Figure \ref{fig:offset}. In the post-processing, thermal emission is calculated every 10$^{\circ}$ of the orbital phase. The phase-curve offsets as a function of wavelength are calculated by fitting a two-term (one and a half orbital period in each term, respectively) sinusoidal function with free amplitudes and offsets to the phase curve at each wavelength, then obtaining the peak offsets at a much finer phase grid. 

At each $T_{\rm eq}$, the phase-curve amplitudes $1-F_{\rm min}/F_{\rm max}$ of the cloud-free models are strongly correlated to molecular spectral features. Smaller amplitudes tend to occur inside molecular windows such as the $J$, $H$ and $K$ bands associated with the H$_2$O at near-IR. TiO and VO features are present at red optical wavelengths, as well as larger amplitudes related to CO opacity near 4.3 $\mu$m. The phase-curve offsets exhibit robust correlations in wavelengths to the phase-curve amplitudes, in which larger offsets appear in wavelengths where amplitudes are small, {\rev the same trend as those found in cloudy models of \cite{parmentier2021}.} These correlations can be intuitively understood as the role of longer radiative timescales in the deeper atmosphere probed by the opacity windows, which permits more efficient heat transport and larger eastward offsets. {\rev We do not find westward phase-curve offsets in the optical wavelengths, which have been shown in phase-curve observations of NGTS-10 Ab \citep{Coulombe2026}. This is because we do not include reflection components in our post-processing; we expect this effect to be small in our models because cloud coverages are relatively limited on the daysides. }

In general, cloudy models have larger amplitudes and smaller phase offsets than cloud-free models, qualitatively consistent with previous work \citep{roman2019,roman2021,parmentier2021}. The differences between cloudy and cloud-free runs are prominent at lower \teq, and become smaller as $T_{\rm eq}$ increases, because the total cloud mass in the atmosphere decreases with increasing \teq. The very slight change in the equatorial jet speeds (Figure \ref{fig:jets}) at a certain equilibrium temperature is likely not responsible for the phase-curve offset differences from the point of view of pure heat advection by the zonal jets. To change the phase offsets by pure jet advection, a much larger variation in wind speed is likely needed (e.g., \citealt{zhang2017}); instead, changes in global wave structures are responsible, as will be discussed further below. 

\begin{figure}
    \begin{center}
    \includegraphics[width=0.5\textwidth]{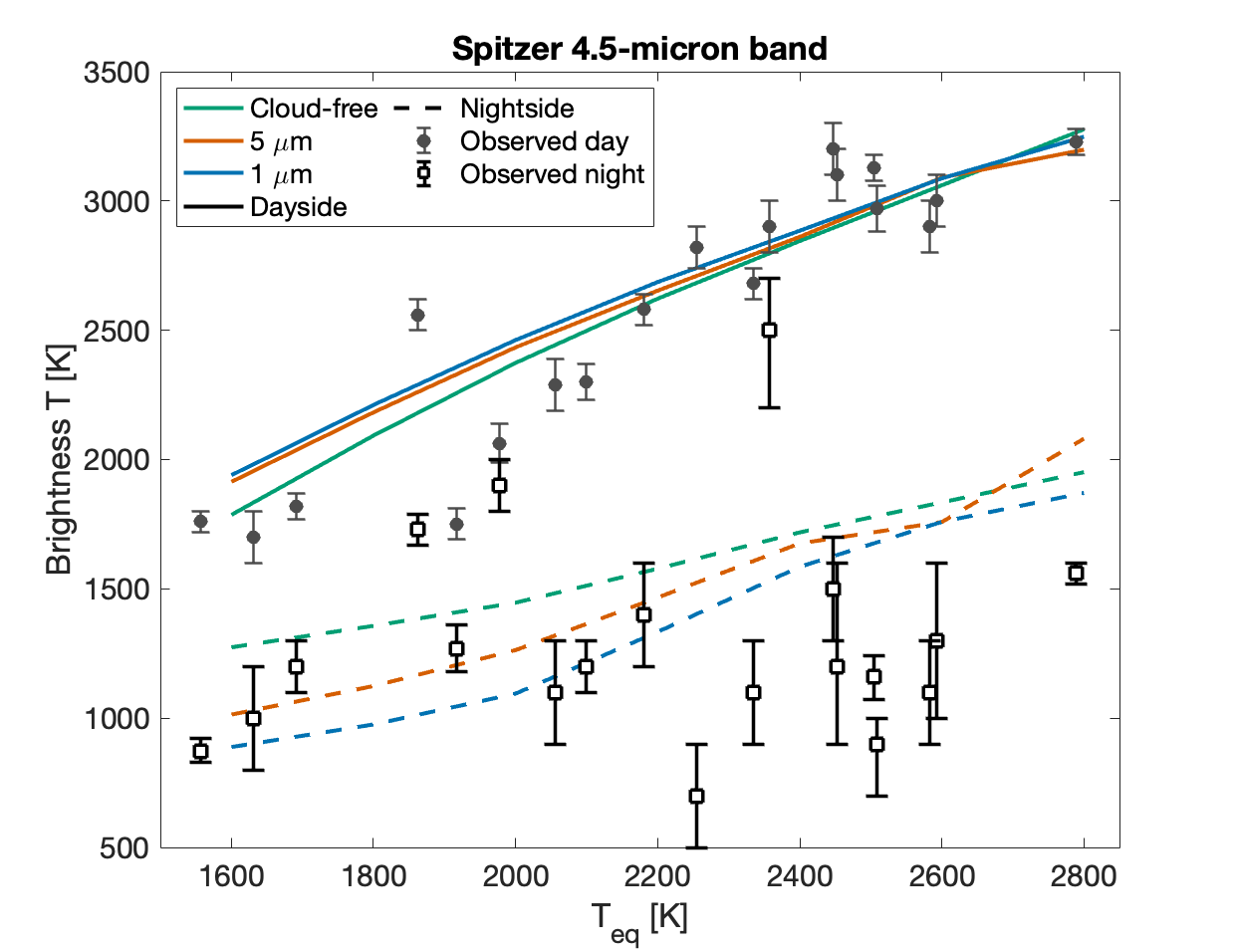}
    \caption{Dayside and nightside brightness temperatures as a function of equilibrium temperature from drag-free GCMs are shown as solid and dashed lines, respectively. {\rev Model colors follow Figure~\ref{fig:offset}. Filled gray circles and open squares show the observed Spitzer 4.5-$\mu$m dayside and nightside temperatures, respectively, with error bars from \cite{dang2025}.} The brightness temperature $T_b$ derived from models is defined such that $\int_{\lambda = 0}^{\infty} \pi B_{\lambda}(T_b) \phi(\lambda) d\lambda = \int_{\lambda = 0}^{\infty} F_{\lambda} \phi(\lambda) d\lambda$, where $F_{\lambda}$ is the model thermal spectrum, and $\phi(\lambda)$ is the instrumental filter profile. 
    }
    \label{fig:day-night}
    \end{center}
\end{figure}

One of the key findings from the cloudy GCMs is that the nightside brightness temperature has a steeper trend as a function of $T_{\rm eq}$ than that of cloud-free GCMs, as shown in Figure \ref{fig:day-night}. The nightside brightness temperatures in the Spitzer 4.5 $\mu$m band of the cloudy model are sufficiently low to explain the observed values at \teq$\lesssim 2200$ K. However, all models overestimate the nightside brightness temperature at \teq$\gtrsim 2200$, regardless of the cloud properties. The nightside brightness temperature of cloudy models increases even faster than that of the cloud-free models as \teq increases. This results from the progressing cloud clearing at the nightside equator.  Clouds alone cannot explain the observed low nightside brightness temperatures by Spitzer, as well as their shallow (or nearly constant) trend as a function of equilibrium temperature.

\subsection{Phase-curve offsets: dynamical and geometric effects}
\begin{figure*}
    \begin{center}
    \includegraphics[width=0.9\textwidth]{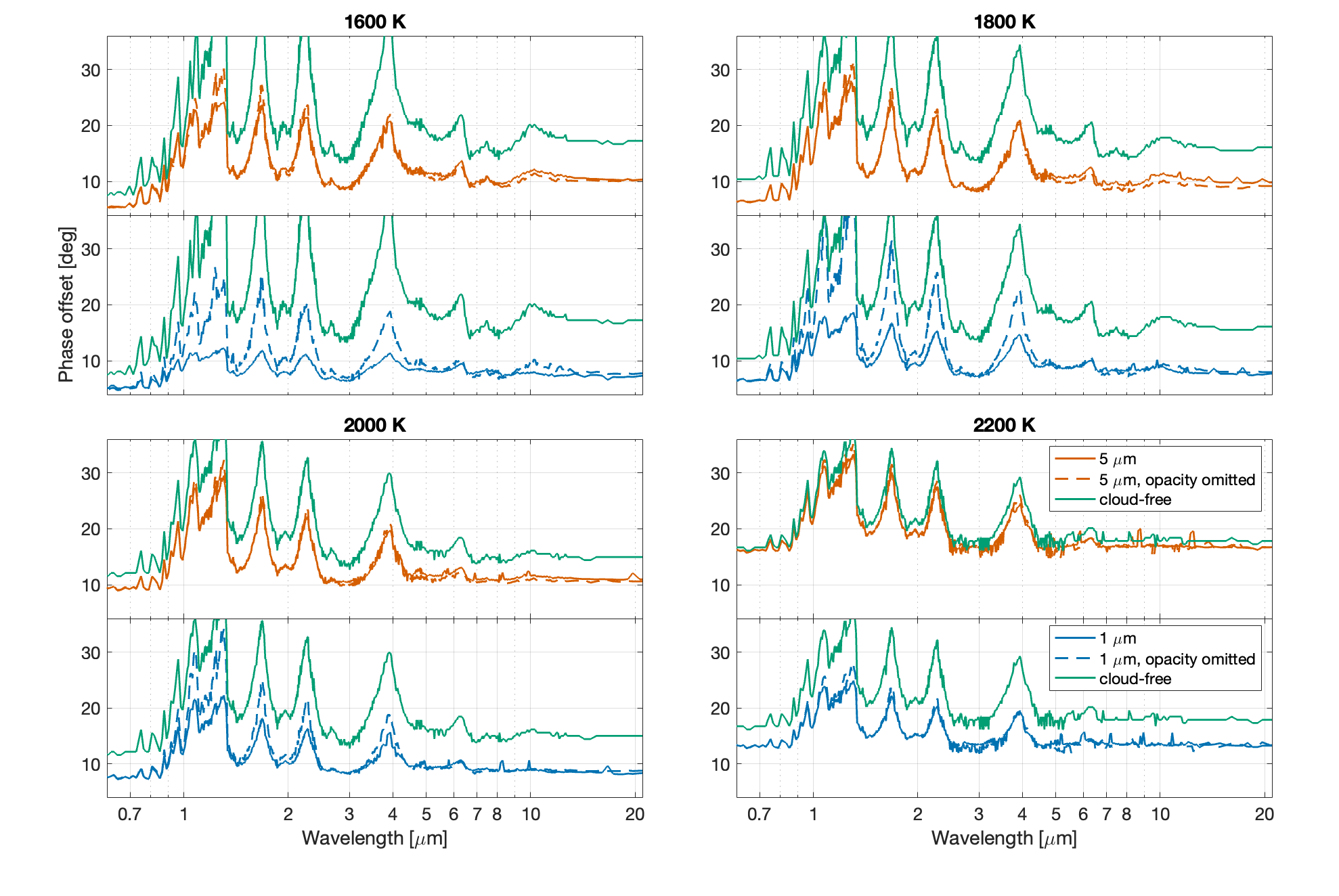}
    \caption{Eastward offsets of thermal phase curves as a function of wavelength for a subset of drag-free models. {\rev The green curve is the cloud-free reference. For the $5~\mu$m (vermilion) and $1~\mu$m (blue) models, solid and dashed curves show offsets calculated from the same GCM temperature structure with cloud opacity included and omitted, respectively.}}
    \label{fig:offset_nc}
    \end{center}
\end{figure*}

Phase-curve offsets are affected by both the spatial temperature structures and cloud distributions. The former relates to the circulation in which temperatures are shaped by various wind components of the circulation \citep{showman2013b,lewis2022}. The latter here is specifically referred to as that caused by the much lower brightness temperatures associated with cloud-tops on the nightside. {\rev The hemispheric averaging that happens in phase curves, together with the stronger brightness contrast between day and night sides, reduce the thermal phase-curve offsets \citep{parmentier2021}.} Having clouds and their radiative feedbacks to the thermal structure and circulation, both effects are at play to modulate the phase-curve offsets. We term the former a ``dynamical effect" and the latter a ``geometric effect". These effects can be diagnosed and clarified in our models by calculating post-processed phase curves in which we artificially remove cloud opacities in the cloudy GCMs. The new sets of phase-curve offsets are then compared to the original values, as shown in Figure \ref{fig:offset_nc} for cases with $r_0=5$ and $1~\mu{\rm m}$ and $T_{\rm eq}=1600$ to 2200 K. Phase-curve offsets of cloud-free GCMs are also displayed. Cases with temperatures higher than 2200 K show limited differences between the two kinds of phase-curve offsets.

All cloudy cases, either with $r_0=1~\mu{\rm m}$ or $r_0=5~\mu{\rm m}$, have significant dynamical effects on the phase-curve offsets, and this is observed from the fact that their phase-curve offsets obtained when clouds are artificially removed in the post-processing are generally much smaller than those of cloud-free GCMs.  At all temperatures shown in Figure \ref{fig:offset_nc}, ``no-cloud-in-post-processing" phase-curve offsets in cases with a mean cloud size of $r_0=5~\mu{\rm m}$ are very similar to the original phase-curve offsets (top panels at each $T_{\rm eq}$). The role of clouds in modulating the phase-curve offsets of cases with larger cloud particles of $r_0=5~\mu{\rm m}$ is primarily due to the ``dynamical effect,'' which alters the circulation and temperature structures through radiative feedbacks. Cases with a mean cloud particle size of $r_0=1~\mu{\rm m}$ show significantly different phase-curve offsets between the two types of phase-curve calculations (bottom panels at each $T_{\rm eq}$), suggesting the ``geometric effect" is also profound in the small-particle cases. Clouds with $r_0=1~\mu{\rm m}$ have longer gravitational sedimentation timescales and are easier to transport to higher altitudes. This results in larger contrasts in cloud-top brightness between the dayside and the nightside than in cases with $r_0=5~\mu{\rm m}$, thereby promoting the geometry effect and reducing the thermal phase-curve offsets.


Why can cloud radiative effects dynamically reduce the amplitudes of phase-curve offsets? {\rev Clouds' greenhouse effects could elevate the dayside temperature and decrease its radiative timescale, potentially leading to smaller phase-curve offsets associated with jet advection \citep{parmentier2016}. However, even though this may certainly have an effect, it is likely not the dominant effect. This is already evident from our results. If increasing the dayside temperature alone were the dominant effect, the hotter cloud-free models should have easily shown smaller offsets than the cloudy, cooler models; however, this is not the case in Figure \ref{fig:offset}. For example, the cloud-free case with T$_{\rm eq}$=2400 K has a much hotter dayside than that in the cloudy cases of T$_{\rm eq}$=1600 K (Figure \ref{fig:maps}); yet the former has larger phase-curve offsets.}

The detailed dynamical mechanisms are yet unclear, but we provide a speculation. Clouds introduce substantial thermal opacity in addition to the gas opacity to the atmosphere. As clouds are relatively well mixed, the thermal opacity in the upper atmospheres increases relative to the cloud-free conditions, effectively decreasing the thermal photospheric pressures, especially on the nightside. This reduces the overall thermal radiative timescale \citep{showman2002}, holding the spatial thermal structure closer to the radiative equilibrium against transports associated with atmospheric wave structures \citep{showman2011,komacek2016,Hammond:2018aa} and jet-driven temperature advections \citep[e.g.,][]{tsai2014,zhang2017}, both of which are sensitive to the thermal radiative timescale. This is analogous to the decrease of phase-curve offsets due to increasing atmospheric metallicity, in which the thermal photospheric pressures also decrease, as shown by recent GCM experiments \citep[e.g.,][]{tan2024,roth2024}. {\rev Although the change mostly happens on the nightside, the wave and jet dynamics can be affected by it because it depends on the day-to-night heating and cooling contrast \citep{showman2011}.}

{\rev We apply the diagnostic technique of exoplanet circulation used in \cite{hammond2021} to decompose the atmospheric circulation into the divergent and rotational components, based on which the temperature field can be further decomposed based on the momentum equation \citep{lewis2022}. Doing so, it might be easier to visualize what temperature structure corresponds to the reduction in phase-curve offsets. The total temperature field of the GCM can be decomposed into two components: the rotational and divergent components, obtained from the hydrostatic relations. Details of this diagnose and interpretations are refereed to \cite{lewis2022}.}

\begin{figure*}
    \begin{center}
    \includegraphics[width=0.9\textwidth]{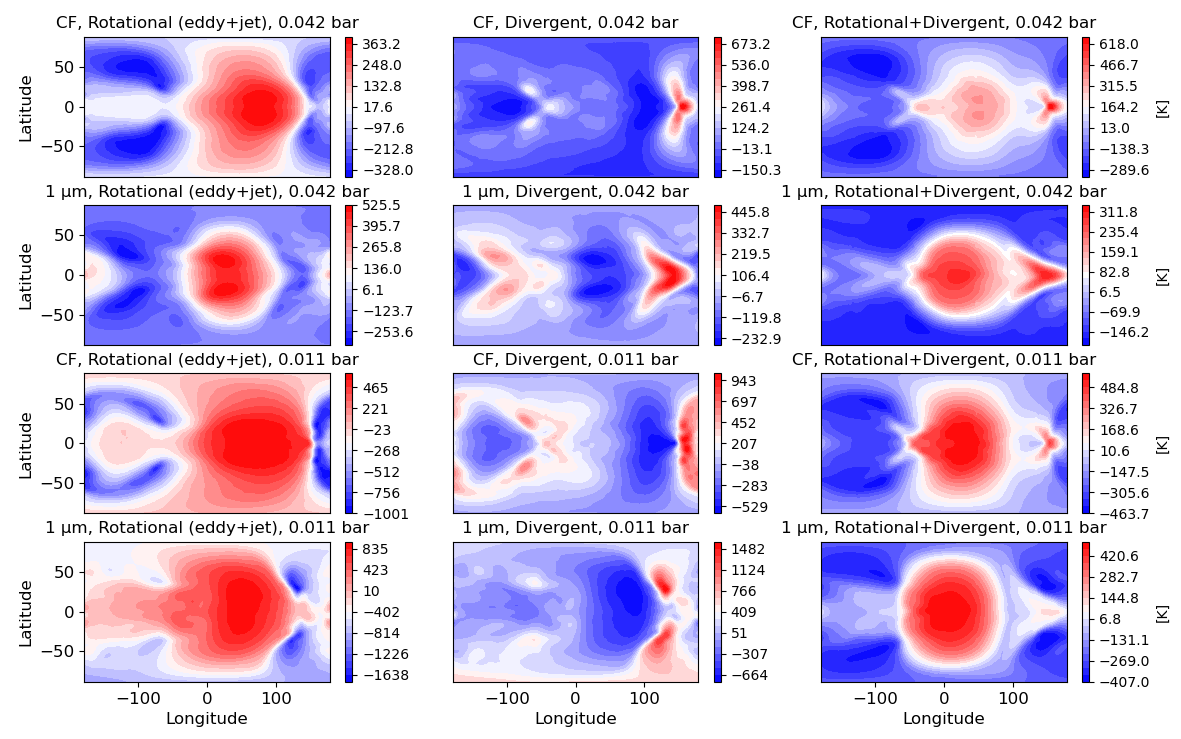}
    \caption{Temperature components corresponding to the balances in the rotational and divergent components of the circulation and their sum for cases of $T_{\rm eq}=1600$ K with cloud-free (CF) condition and cloudy condition with a reference size $r_0=1~\mu$m. Maps at two representative pressures of 0.042 and 0.011 bars are shown. }
    \label{fig:ro_div}
    \end{center}
\end{figure*}

Temperature maps associated with the rotational and divergent components for cloud-free and $r_0=1~\mu$m cloudy cases with $T_{\rm eq}=1600$ K are shown in the left and middle columns of Figure \ref{fig:ro_div}. The specific gas constant $R$ in the GCM can change as a function of pressure and temperature, given our hydrogen dissociation treatment; however, photospheres with $T_{\rm eq}=1600$ K have minor changes of $R$, which was treated as a constant.  Rotational components, intuitively speaking, mostly contributed by the zonal jets and standing Rossby waves in the off-equatorial regions, are mainly responsible for the eastward temperature shifts. The divergent components, in general, have the ``wrong" zonal distribution to explain the eastward phase offsets; therefore, they do not directly contribute to the phase offsets. Indeed, a visual inspection of the sum of rotational and divergent components shown in the right column of Figure \ref{fig:ro_div} suggests that the rotational components matter for the eastward offsets.  The components in the cloudy cases, shown at two representative pressure layers near the thermal photosphere, are all less shifted than those in the cloud-free cases. This matches our expectation that the overall lower thermal radiative timescale of the cloudy cases might help reduce the wave shifts against the zonal jets. 

\section{GCMs with Drag Implementations}
\label{sec:drag}
{\rev In the drag-free models, equatorial jets maintain warm, relatively clear regions on the nightside at $T_{\rm eq}\gtrsim2200$~K (Figures~\ref{fig:maps} and \ref{fig:day-night}). Here we examine how damping the horizontal flow changes nightside heat transport and cloud coverage, and whether the resulting circulation can account for both the nightside brightness temperatures and the winds inferred at the limb.}

\subsection{Drag Implementations}
We parameterize the dissipation of large-scale flow as ``drags" on the horizontal wind. Our purpose is to suppress the global wave patterns and the strong zonal jets, rather than realistically simulate the full large-scale dynamics under realistic MHD conditions. We choose two idealized drag schemes as follows. 

The first drag scheme is a linear Rayleigh drag applied in the angular momentum equation, $-\mathbf{u}/\tau_{\rm drag}$ where $\mathbf{u}$ is the horizontal wind, with a constant drag timescale $\tau_{\rm drag}$ that is held fixed in all space and time. The drag timescale is assumed to be $10^4$ s, as in our precursor work of \cite{tan2019uhj,tan2024}. {\rev This timescale is shorter than the rotation period of a typical hot Jupiter,} $\sim 10^5$ s. By timescale considerations, this drag starts disrupting the complete formation of the large-scale {\rev Matsuno-Gill} waves \citep{showman2011,komacek2016} and suppresses the strong equatorial super-rotating jet, therefore qualitatively and quantitatively influences the cloud distributions.

In another set of simulations, we also apply the linear magnetic drag parameterization, same form as above but in which $\tau_{\rm drag}(T,P,\chi)$ depends on the pressure and temperature at a given composition of minor (ionizing) species $\chi$ (\citealt{perna2010}, see also the original development by \citealt{liu2008} for the Ohmic dissipation associated with Jovian zonal jets).  Note that we adopt the version of magnetic drag using an order-of-magnitude estimation of the induced currents based on only local information \citep{Menou2012,rauscher2013}, rather than a detailed reconstruction based on the global wind and magnetic fields (essentially means that local currents are determined by non-local information). Note that there could be significant differences in both the spatial drag distribution and strengths between the two treatments \citep{Wazny2025}. Unlike previous work that applied the drag only on the zonal direction \citep[e.g.,][]{perna2010,rauscher2013}, we apply the drag to both the zonal and meridional directions of the horizontal winds. More rigorous forms of the magnetic drag in all directions, with also assumptions of small induced magnetic fields and order-of-magnitude estimation of local currents, have been proposed by recent work \citep{Christie2025,Blocker2026}. The magnetic drag applied in this work does not account for the above non-local reconstruction or rigorous 3D drag forms, because we focus on understanding the effects of clouds and keep the drag implementations idealized.  We assume a dipolar background magnetic field of 10 G, within the range considered in previous work \citep[e.g.,][]{rogers2014komacek,beltz2022}. The ionization fraction at each location and time of the atmosphere is obtained from the equilibrium chemical calculations of \cite{visscher2006} assuming $1\times$solar metallicity.

{\rev At a given pressure, the increase in thermal ionization with temperature generally gives shorter local drag timescales on the irradiated dayside and longer timescales on the cooler nightside. This prescription therefore allows the damping to vary both around a planet and across the irradiation sequence. We apply $\tau_{\rm drag}=\max(\tau_{\rm local},\tau_{\min})$, where $\tau_{\rm local}$ is the timescale from the magnetic prescription. The floor limits the maximum damping rate in the regime where the assumption of a small induced field becomes uncertain \citep{liu2008}. Its value is a modeling choice. We adopt $\tau_{\min}=10^2$~s for the magnetic-drag models throughout this work. For each cloud treatment we also retain a sequence with $\tau_{\min}=10^4$~s to test the sensitivity to stronger damping where the local timescale falls below $10^4$~s. We identify this comparison sequence explicitly by its floor.}

Overall, regardless of the details of drag implementations, our purpose is to suppress wave and jet formation. Indeed, full global MHD simulations of hot Jupiters' atmospheres exhibit dissipation of the overall wind speed and suppression of the hydrodynamic Matsuno-Gill waves and the strong equatorial jets \citep{rogers2014,rogers2014komacek,rogers2017b,boning2025}. 

\subsection{Results}
\begin{figure*}
    \begin{center}
    \includegraphics[width=1\textwidth]{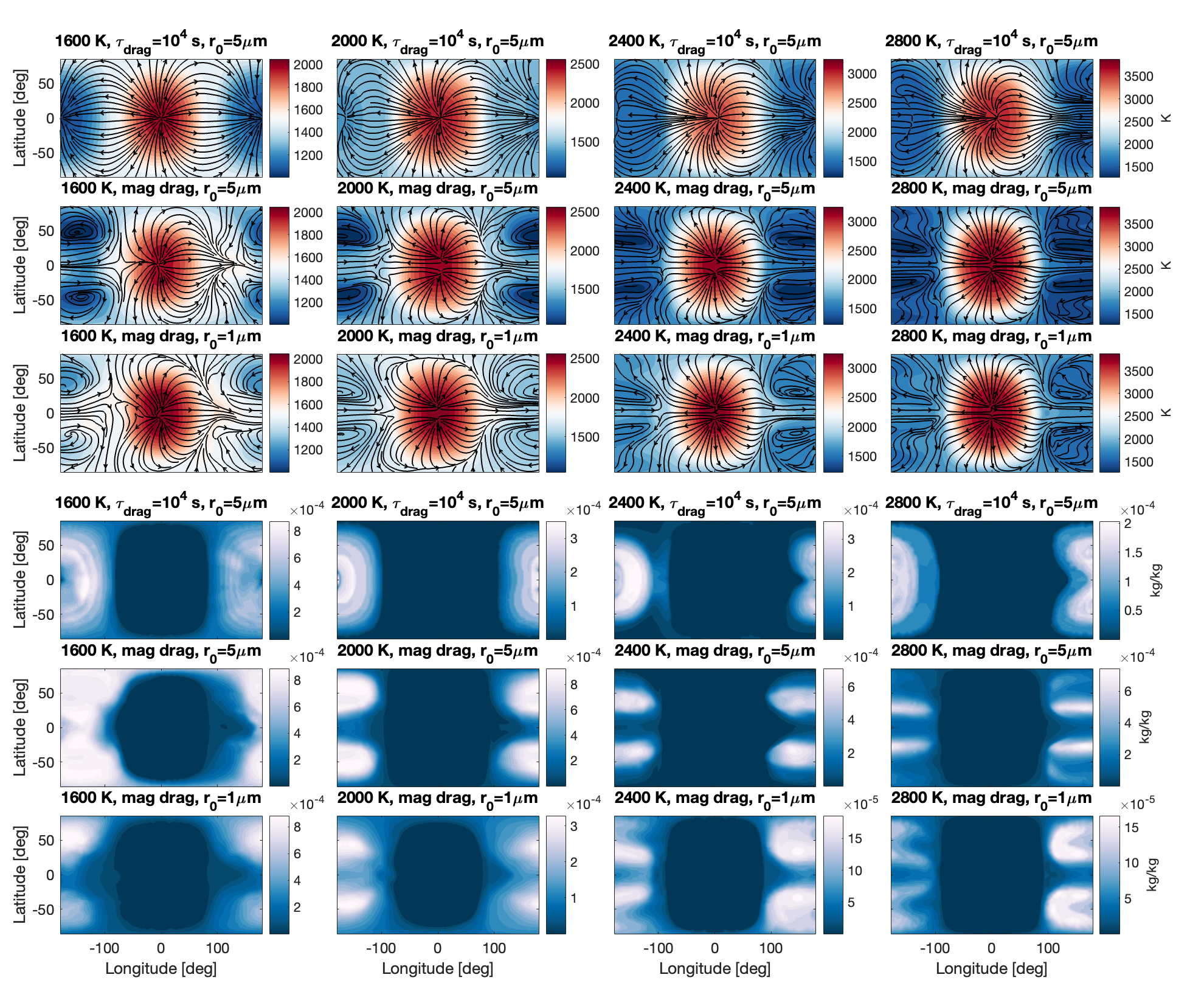}
    \caption{Similar to Figure \ref{fig:maps} but showing temperature maps at 32 mbar and cloud distribution for cloudy models with $r_0=5~\mu{\rm m}$ and a uniform drag timescale $\tau_{\rm drag}=10^4$ s in the first and fourth row;  those for $r_0=5~\mu{\rm m}$ and magnetic drag {\rev ($\tau_{\min}=10^2$~s)} are in the second and fifth row; and those for $r_0=1~\mu{\rm m}$ and magnetic drag are in the third and sixth row. }
    \label{fig:maps_drag}
    \end{center}
\end{figure*}

The drag has a direct impact on the qualitative circulation pattern, as shown in temperature maps, wind directions and horizontal cloud distributions in Figure \ref{fig:maps_drag}. In cases with a constant drag timescale $\tau_{\rm drag}=10^4$ s, the lowest temperature at which the rotation period is $ 3.8\times10^5~{\rm s}\gg \tau_{\rm drag}$ shows a complete suppression of waves and zonal jets. A day-to-night overturning flow is a result of the balance between the pressure gradient and the drag \citep{perezbecker2013,komacek2016}. Therefore, the temperature and cloud distributions are, to the first order, symmetrical about the substellar points.  As the rotation period decreases for higher temperatures, rotational effects develop, and the tropical wave patterns gradually emerge, especially in the highest temperature case with $T_{\rm eq}=2800$ K, although zonal jets are still suppressed. A consequence of the waves is that the temperature and cloud distribution exhibit east-west asymmetry. 
Critically, the suppression of the zonal jet formation and heat transport leads to a larger cloud coverage on the nightside. This is especially obvious in the cases with a constant drag timescale.

In the ``magnetic drag" treatment, the ionization fraction is sensitive to temperature and pressure, resulting in spatially dependent drag timescales. The resulting circulation pattern is no longer simply a global day-to-night pattern as in the case of a uniform drag timescale. On the cooler nightside, where the drag is very weak, stationary vortices at mid-latitudes persist, similar to the drag-free cases. On the dayside, especially near the substellar point, the flow symmetrically diverges, similar to the case of a uniform drag timescale. Not surprisingly, the zonal jets are suppressed even when the strong damping occurs mostly on the dayside, similar to previous GCM studies with parameterized magnetic drags \citep[e.g.,][]{beltz2022,kennedy2025}. A subtle feature is that the nightside persisting Rossby-wave-like vortices move closer to the equator compared to the drag-free cases, and similar patterns were also found in previous GCMs using ``magnetic drag" formalism, albeit only on the zonal direction \citep{beltz2022,kennedy2025}. The dynamical mechanisms responsible for nightside vortices being closer to the equator are unclear. The cold regions with thicker clouds associated with the vortices move closer to the equator, where it is observationally biased in phase curves, thereby reducing brightness temperatures of the nightside. 

\begin{figure*}
    \centering
    \includegraphics[width=0.9\textwidth]{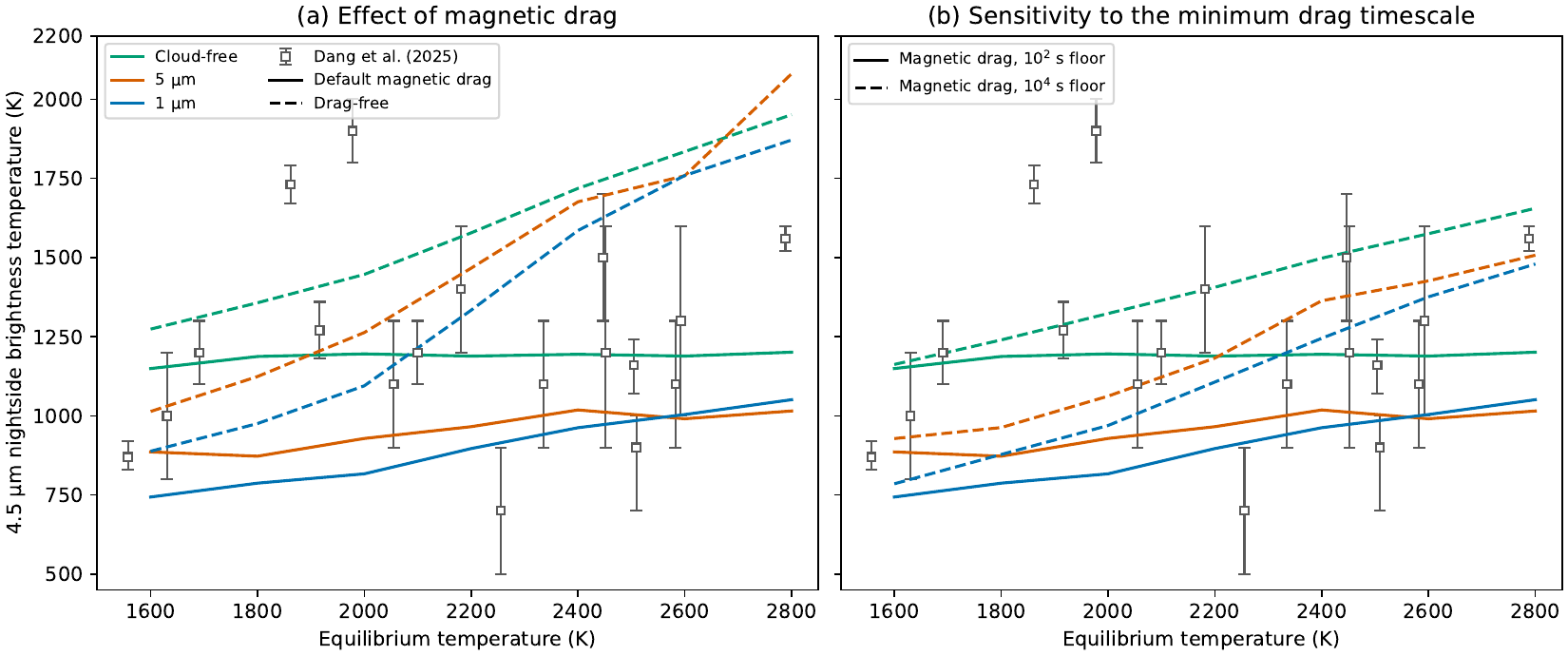}
    \caption{{\rev Spitzer 4.5-$\mu$m nightside brightness temperatures versus equilibrium temperature. Green, vermilion, and blue denote the cloud-free, 5-$\mu$m, and 1-$\mu$m sequences, respectively. Solid curves show the default magnetic-drag models ($\tau_{\min}=10^2$~s) in both panels. Dashed curves show the drag-free models on the left and the magnetic-drag models with $\tau_{\min}=10^4$~s on the right. Squares show the measurements of \cite{dang2025}.}}
    \label{fig:T_night}
\end{figure*}

{\rev Figure~\ref{fig:T_night} shows that magnetic drag changes both the level of the nightside brightness temperature and its dependence on irradiation. At 4.5~$\mu$m, the cloud-free magnetic-drag sequence stays within $\sim$1190-1200~K from $T_{\rm eq}=1800$ to 2800~K. Clouds lower the nightside brightness further. Across the full grid ($T_{\rm eq}=1600$-2800~K), the brightness temperature rises from $\sim$890 to 1020~K for $r_0=5~\mu$m and from $sim$750 to 1050~K for $r_0=1~\mu$m. Thus a nearly constant trend is possible even in a cloud-free atmosphere, while the cloudy sequences retain a dependence on particle size and irradiation. The corresponding 3.6-$\mu$m sequences also flatten when the minimum drag timescale is at $10^2$ s.}

\begin{figure*}
    \begin{center}
    \includegraphics[width=1\textwidth]{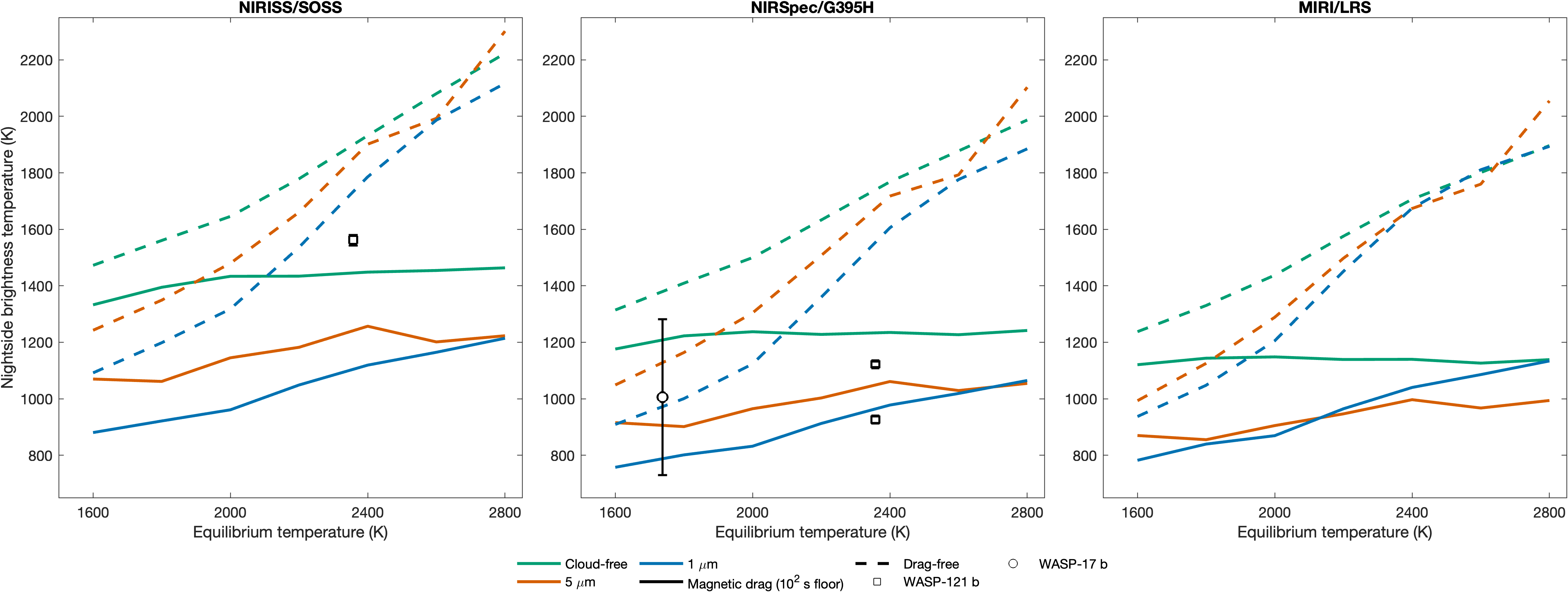}
    \caption{{\rev Nightside brightness temperatures in the JWST NIRISS/SOSS, NIRSpec/G395H, and MIRI/LRS bands. Green, vermilion, and blue represent the cloud-free, 5-$\mu$m, and 1-$\mu$m sequences. Solid curves show magnetic drag ($\tau_{\min}=10^2$~s) and dashed curves show drag-free models. Open squares denote WASP-121 b and the open circle denotes WASP-17 b.} The data point in the panel for NIRISS/SOSS is the nightside brightness temperature of WASP-121 b derived in \cite{Splinter2025}. In the panel for NIRSpec/G395H, the data point near \teq$\sim1750$ K is the nightside brightness temperature of WASP-17 b derived in \cite{Lustig-Yaeger2025}; and the two data points near \teq$\sim2350$ K are those of WASP-121 b from \cite{Mikal-Evans2023} (the two points correspond to NRS1 and NRS2 bands, respectively). }
    \label{fig:T_night_other}
    \end{center}
\end{figure*}

{\rev The uniform-drag models provide a control for the effect of damping the circulation with the same timescale on both hemispheres (Figure~\ref{fig:maps_drag}). Although a uniform $\tau_{\rm drag}=10^4$~s suppresses the equatorial jet, the cloud-free 4.5-$\mu$m nightside brightness temperature still rises from $\sim$1050 to 1650~K across our grid, compared with $\sim$1150 to 1200~K with the default magnetic drag. We therefore retain the uniform-drag comparison in the circulation maps and focus Figures~\ref{fig:T_night} and~\ref{fig:T_night_other} on how clouds and the spatially varying magnetic drag affect the nightside emission.}

{\rev With the higher drag floor of $10^4$~s, the 4.5-$\mu$m nightside brightness temperature instead rises by about 490, 580, and 700~K across the grid for the clear, 5-$\mu$m, and 1-$\mu$m sequences, respectively. The differences between the two drag prescriptions grow toward high irradiation, as expected if increasing ionization allows more effective damping of heat transport in the hotter models. Lowering the drag floor brings the modeled slopes closer to the shallow trend inferred from Spitzer phase curves \citep{beatty2019,bell2021,dang2025}. The predicted brightness temperatures also depend on the wavelength range sampled, as illustrated for three JWST bands in Figure~\ref{fig:T_night_other}. At $T_{\rm eq}=2800$~K, adding 5-$\mu$m clouds to the magnetic-drag models lowers the nightside brightness temperature by approximately 240~K in SOSS, 190~K in G395H, and 150~K in MIRI/LRS. Section~\ref{sec:los_velocity} examines how the same change in drag affects winds at the limb.}

\subsection{Line-of-sight Atmospheric Velocities}
\label{sec:los_velocity}

{\rev High-resolution transmission spectroscopy provides a complementary probe of atmospheric circulation through the Doppler shifts of absorption lines. In HD~209458~b, \cite{snellen2010} measured a CO blueshift of about 2~km~s$^{-1}$, consistent with flow from the dayside to the nightside. Using phase-resolved sodium absorption in HD~189733~b, \cite{louden2015} inferred a redshift of about 2~km~s$^{-1}$ on the leading limb and a blueshift of about 5~km~s$^{-1}$ on the trailing limb, including the contribution from planetary rotation. In the UHJ WASP-76~b, the blueshift of neutral iron absorption increases during transit, which \cite{ehrenreich2020} attributed to atmospheric motion combined with an asymmetry in gaseous iron abundance between the limbs. Because different species and transitions probe different atmospheric regions, these measurements can also constrain the vertical structure and morning--evening asymmetry of the flow. For example, \cite{seidel2025} inferred a vertically structured circulation on WASP-121 b using iron, sodium, and hydrogen absorption, while \cite{seidel2026} found a systematic decrease in the iron-line day-to-night wind speeds across a sample of seven UHJs with increasing irradiation. These observations motivate examining the pressure-dependent line-of-sight (LOS) velocities in our model grid, in addition to the thermal phase curves.}

\begin{figure*}
    \centering
    \includegraphics[width=\textwidth,height=0.86\textheight,keepaspectratio]{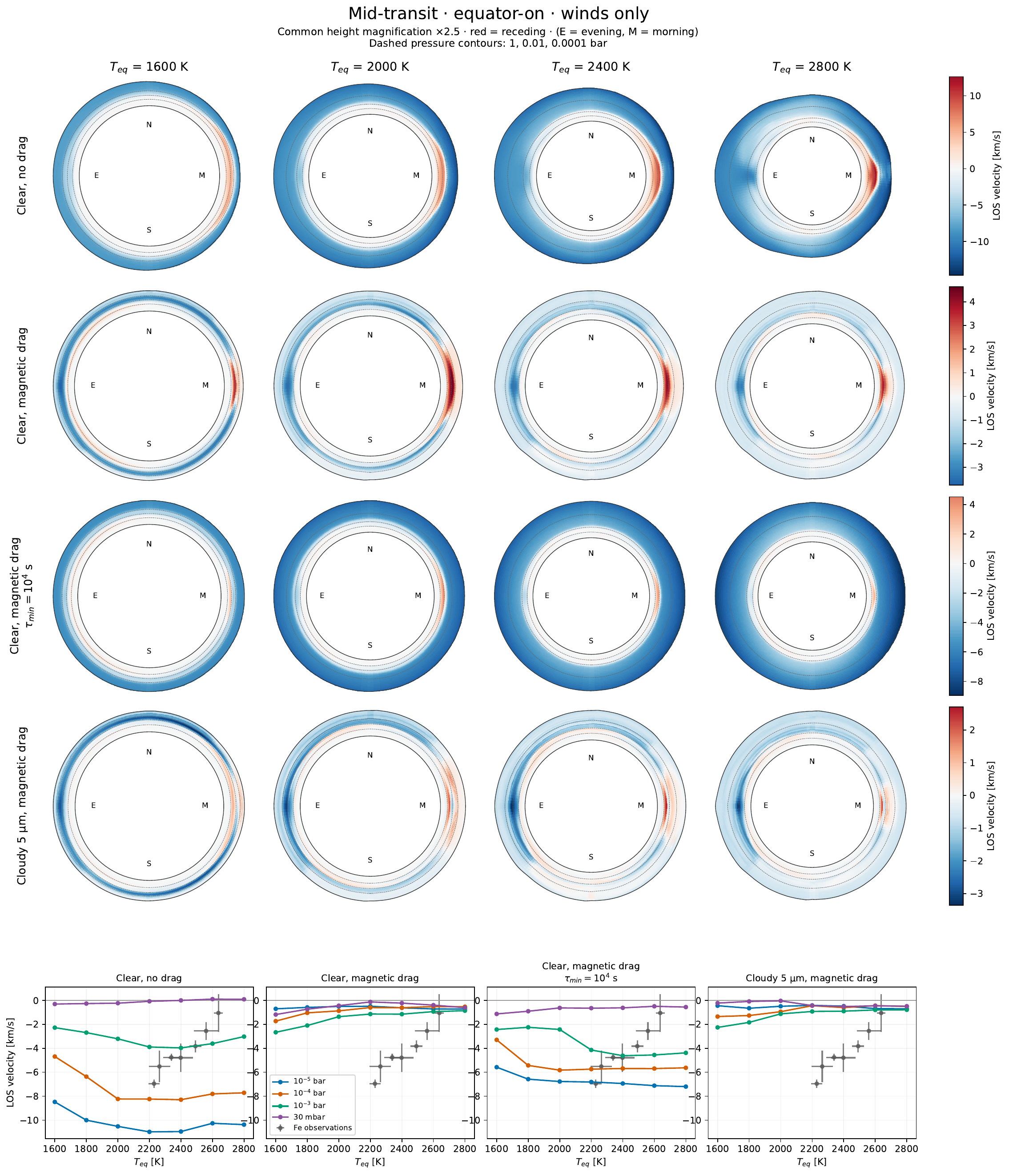}
    \caption{{\rev Line-of-sight winds at mid-transit, excluding rotation. Maps show time-averaged winds within $\pm5^\circ$ of the limb for the four model families and equilibrium temperatures indicated. Each row has its own color scale. Blue indicates approaching flow. Heights are magnified $2.5\times$, and dashed contours mark 1, 0.01, and $10^{-4}$~bar. Bottom panels compare signed azimuthal means at four pressures with the iron-line measurements of \cite{seidel2026}, shown with blueshifts negative. Model means use geometric weighting (Section~\ref{sec:los_velocity}).}}
    \label{fig:los_velocity}
\end{figure*}

{\rev Figure~\ref{fig:los_velocity} shows the LOS velocities for an equator-on observer at mid-transit. We project the time-averaged GCM winds onto the observer's direction and average over a $\pm5^{\circ}$ belt around the apparent limb, with solid-angle weighting across the belt. Negative velocities denote approaching motion, corresponding to blueshifts. Solid-body rotation is excluded to isolate the atmospheric circulation. The upper four rows show the dependence on position around the limb and pressure for cloud-free models without magnetic drag, cloud-free models with magnetic drag, a comparison magnetic-drag sequence with a minimum drag timescale of $10^4$~s, and cloudy magnetic-drag models with $r_0=5~\mu$m. The two magnetic-drag sequences are distinguished here because the imposed minimum timescale substantially affects the upper-atmosphere circulation. Hydrostatic atmospheric heights are magnified by a common factor of 2.5 to display the vertical structure; each row has its own velocity color scale.}

{\rev Without magnetic drag, the deeper equatorial flow produces blueshifts on the evening limb and redshifts on the morning limb, as expected for an eastward jet. At lower pressures, approaching motion extends over much of both limbs, indicating a strong day-to-night component. These distinct signatures illustrate why a net Doppler shift does not directly measure the maximum jet speed: opposite LOS velocities associated with zonal flow can cancel when averaged around the limb, whereas day-to-night flow contributes approaching motion on both limbs at transit. The maps retain the spatial information lost in such an average, including the pressure dependence of the morning--evening asymmetry.}

{\rev Magnetic drag substantially reduces the approaching velocities in the upper atmosphere. In the second row, the extended regions of strong blueshift seen in the drag-free models largely disappear, although localized redshifted and blueshifted flows remain near the equatorial limbs. The cloudy models with the same drag treatment exhibit similarly weak upper-atmosphere velocities, with differences in their spatial distribution and vertical structure arising from the cloud-modified thermal forcing. In contrast, the sequence with $\tau_{\min}=10^4$~s retains substantial approaching motion over most of the upper limb. A drag prescription that suppresses the equatorial jet therefore need not suppress the high-altitude day-to-night flow to the same degree. The sensitivity to the drag-timescale floor also cautions against interpreting a particular parameterized-drag model as a unique prediction of magnetic effects.}

{\rev The bottom row summarizes these results using the signed, full-azimuth mean of the belt-averaged LOS velocity at fixed pressure. The drag-free sequence reaches mean velocities of approximately $-8$ to $-11$~km~s$^{-1}$ at $10^{-5}$~bar, whereas the corresponding velocities at $10^{-3}$~bar are approximately $-2$ to $-4$~km~s$^{-1}$. The magnetic-drag sequence with $\tau_{\min}=10^4$~s retains mean blueshifts of several km~s$^{-1}$, including roughly $4$--$7$~km~s$^{-1}$ at $10^{-5}$--$10^{-3}$~bar for the hotter models. The default magnetic-drag sequences ($\tau_{\min}=10^2$~s), both clear and cloudy, instead have mean blueshifts of order $1$~km~s$^{-1}$ or less at the hotter end of the grid. For these selected models and pressures, the difference between the drag treatments is larger than the difference between the clear and cloudy cases with the same drag treatment.}

{\rev The iron-line velocities inferred by \cite{seidel2026} provide a qualitative comparison. Their inferred day-to-night speeds decrease from about 7 to 1~km~s$^{-1}$ across the observed irradiation range. The pressure range sampled by iron depends on the temperature and abundance of neutral iron along the limb, and regions with different velocities can contribute unequally to the line shift. The curves at individual pressures in Figure~\ref{fig:los_velocity} therefore illustrate the vertical variation in the modeled flow. A quantitative comparison would require phase-resolved transmission spectra that account for these variations, cloud opacity, and rotation.}

{\rev The clearest difference among the models occurs at low pressures. At $10^{-5}$--$10^{-3}$~bar the default magnetic drag strongly damps the net approaching flow in the hotter models, both clear and cloudy, whereas imposing a $10^4$~s minimum timescale allows blueshifts of several km~s$^{-1}$ to persist. Thus the same magnetic-drag models that yield a shallow trend in broadband nightside brightness temperatures also predict very weak line-of-sight winds in the upper atmosphere at the limb, inconsistent with the measured trend shown in \cite{seidel2026}. Our magnetic drag models assume a field strength of 10~G, higher than the few gauss inferred by \citet{seidel2026}, so future simulations spanning a range of field strengths, including weaker fields, are needed to assess whether the models can better reproduce the observed line-of-sight velocities. Taken together, these observations can guide future improvements in how we model magnetic interactions in hot-Jupiter atmospheres by constraining both horizontal heat transport and the pressure dependence of wind damping.}

\section{Discussion}
\label{sec:dis}
\subsection{Nightside Spectra may Break the Degeneracy Between Clouds and Drag}
\begin{figure*}
    \begin{center}
    \includegraphics[width=0.7\textwidth]{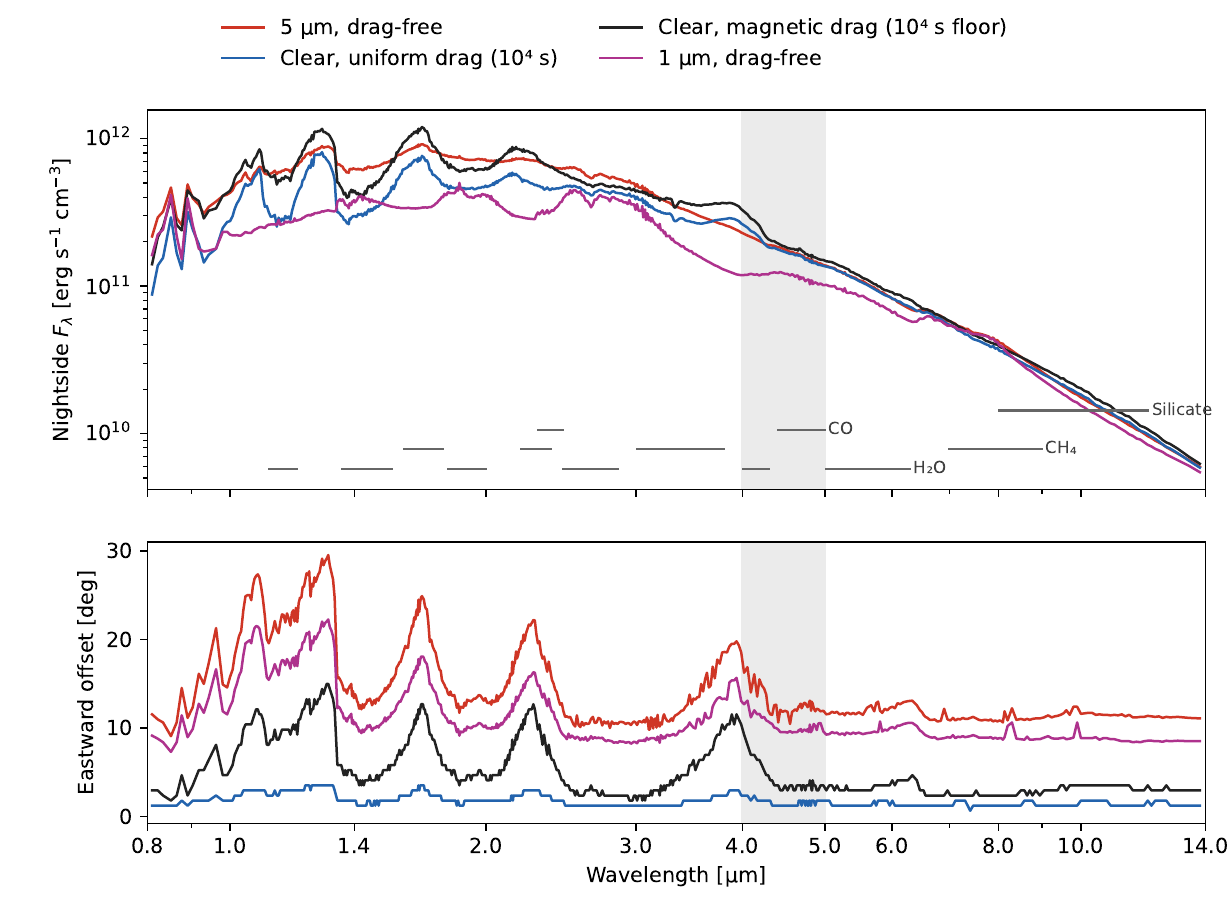}
    \caption{{\rev Nightside emission spectra (top) and eastward thermal phase-curve offsets (bottom) for the same four models at $T_{\rm eq}=2000$~K. The black curve denotes the magnetic-drag comparison model with $\tau_{\min}=10^4$~s. The offsets give the phase of maximum thermal emission from the two-harmonic fit at each wavelength. Horizontal bars mark selected molecular bands and the silicate feature. Gray shading indicates the approximate Spitzer 4.5-$\mu$m band.}}
    \label{fig:spectra2000}
    \end{center}
\end{figure*}

Effects of clouds and large-scale dissipation---``drags" implemented in our models---can be both tuned in the GCMs and result in roughly the same nightside brightness temperatures over a broadband observation, leading to possible degeneracies for interpreting observed phase-curve amplitudes and nightside temperatures. This degeneracy could be broken utilizing the nightside spectra, which can be readily obtained by JWST observations for transiting exoplanets \citep[e.g.,][]{bell2024wasp43b,Evans-Soma2025,Lustig-Yaeger2025}. {\rev We illustrate this degeneracy at $T_{\rm eq}=2000$~K with three models of comparable Spitzer 4.5-$\mu$m nightside brightness (Figure~\ref{fig:spectra2000}): a drag-free cloudy model with $r_0=5~\mu$m (red), a cloud-free model with uniform drag (blue), and the cloud-free magnetic-drag comparison model with $\tau_{\min}=10^4$~s (black).} In addition, a case with drag-free conditions and $r_0=1~\mu$m is shown in magenta lines for a spectral shape comparison, despite having obviously lower fluxes over the Spitzer 4.5-$\mu$m band.

Despite having similar fluxes at longer wavelengths, the nightside spectral amplitudes of the case with drag-free and $r_0=5~\mu$m are substantially weaker than those of cases without clouds but with drag. Clouds are relatively well mixed on the nightside, providing significant cloud continuum opacity to reduce the spectral amplitudes. This has been well understood in self-luminous substellar atmospheres \citep[e.g., reviews by][]{marley2015,zhang2020}. Nightside spectra of two cloud-free cases, one with a uniform drag timescale and the other with a magnetic drag, differ only by an absolute shift rather than by relative spectral amplitudes, to the first order. Other key ingredients being the same, the nightside spectrum is unlikely to be a direct probe of the form of the dissipation on the large-scale flows. {\rev Nightside spectra can thus help distinguish the effects of clouds and atmospheric drag on thermal emission \citep[see also the analysis of WASP-121~b by][]{frazier2026}.}

{\rev The wavelength dependence of the phase offset provides a complementary diagnostic (Figure~\ref{fig:spectra2000}, bottom). Near $1.25~\mu$m, the eastward offset is about $26^\circ$ for the drag-free $5~\mu$m cloud model, compared with $4^\circ$ for the clear uniform-drag model and $12^\circ$ for the clear magnetic-drag comparison model. In the $1.4~\mu$m water band these offsets decrease to about $13^\circ$, $2^\circ$, and $5^\circ$, respectively. Thus models with comparable broadband nightside brightness differ in both the magnitude of their offsets and the contrast between a near-infrared window and an absorption band.}

A more interesting comparison in Figure \ref{fig:spectra2000} is between the case with $r_0=5~\mu$m and $r_0=1~\mu$m. The latter has an overall much smaller flux due to more efficient cloud mixing on the nightside, but exhibits {\it emission} features in the near-IR H$_2$O molecular bands, rather than the {\it absorption} features as in the other cases. This is also in stark contrast to theoretical expectations, which predict that a large fraction of nightsides should generally have temperature structures that decrease with height, giving rise to absorption features. 

\begin{figure*}
    \centering
    \includegraphics[width=0.9\textwidth]{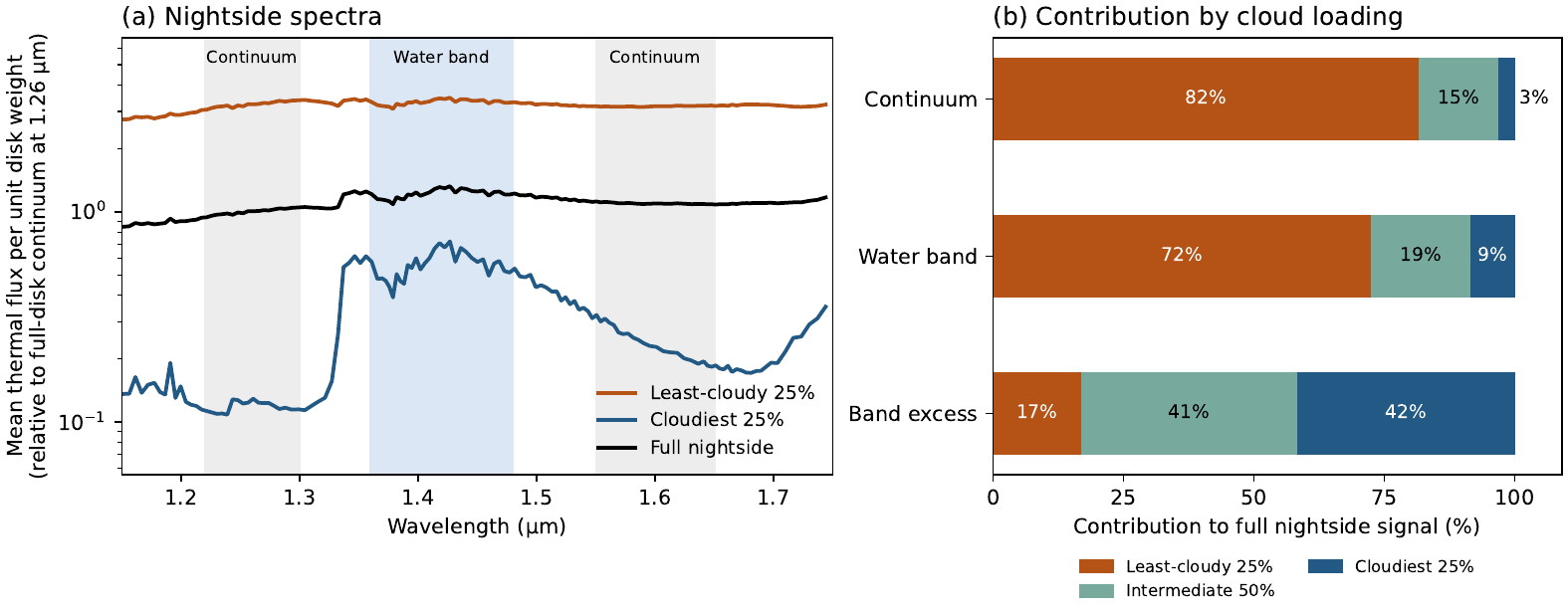}
    \caption{{\rev Origin of the nightside water emission feature in the drag-free model with $T_{\rm eq}=2000$~K and $r_0=1~\mu$m. Columns are ranked by their pressure-weighted mean condensate mass fraction above 0.1~bar. (a) Disk-weighted mean spectra of the lowest and highest quartiles and the full nightside, with a common normalization. Blue shading marks the water band (1.36--1.48~$\mu$m) and gray shading the continuum windows (1.22--1.30 and 1.55--1.65~$\mu$m). (b) Fraction contributed by each group to the continuum, water band, and band excess. For each column we subtract the continuum interpolated between the two windows from the mean water-band flux, then sum these excesses with the disk weights.}}
    \label{fig:cloud_contributions}
\end{figure*}

{\rev To locate the origin of the $1.4~\mu$m water feature, we compare nightside columns grouped by condensate loading (Figure~\ref{fig:cloud_contributions}). The least-cloudy quartile supplies 72\% of the water-band flux but also 82\% of the adjacent continuum, so it contributes only 17\% of the disk-integrated emission excess. The cloudiest quartile is much fainter, yet its water band rises well above its local continuum and accounts for 42\% of the excess. The intermediate half of the columns supplies the remaining 41\%. Thus relatively clear regions dominate the near-infrared flux in this wavelength range, while cloudier regions supply most of the emission contrast. When cloud opacity is removed in post-processing at fixed temperature structure, the disk-integrated band changes from emission to absorption. Thermal scattering can produce such emission features \citep{taylor2021scattering}, although this test removes both cloud absorption and scattering and cannot separate their contributions.} Observationally, nightside emission features have not been found in low-resolution spectra, but emitting H$_2$O lines from the nightside of a UHJ were recently discovered through high-resolution spectroscopy during the primary transit of WASP-33b \citep{yang2024}. Lastly, the case with $r_0=1~\mu$m shows a recognizable silicate absorption feature near 10 $\mu$m.

\subsection{Low Bond albedos and Overestimations by Energy Balance Models}
\begin{figure}
    \begin{center}
    \includegraphics[width=0.5\textwidth]{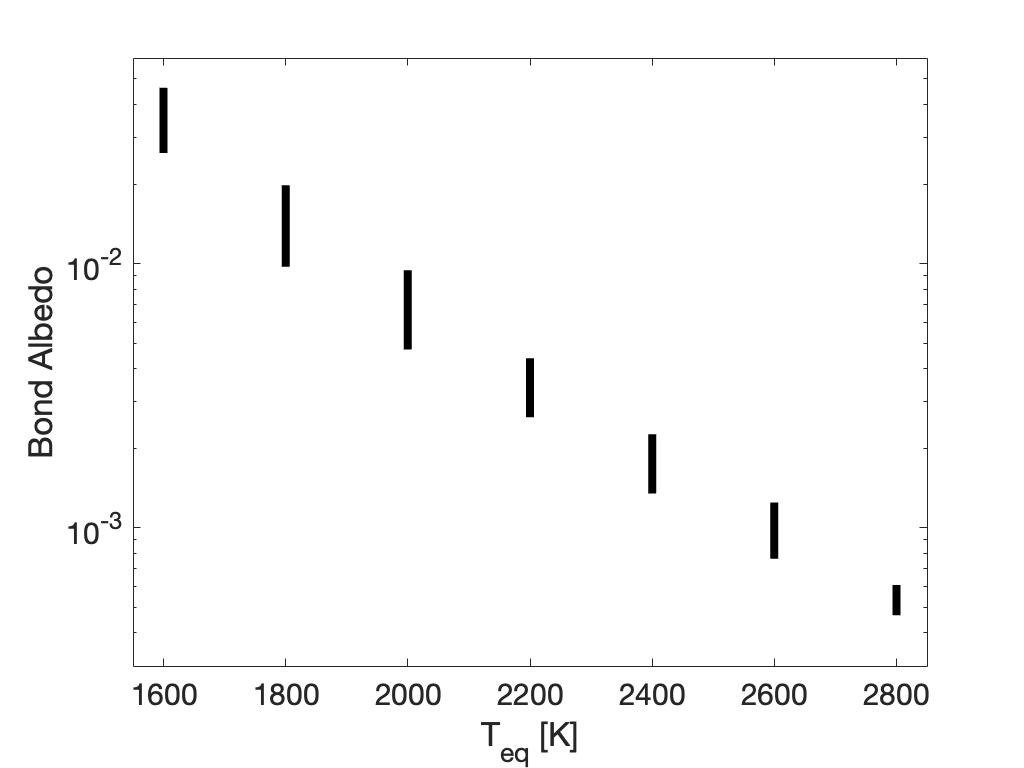}
    \caption{Ranges of Bond albedos for each $T_{\rm eq}$ {\rev diagnosed from the cloud-free and two cloudy drag-free GCM sequences}. For each $T_{\rm eq}$, the highest $A_b$ is from the case with $r_0=1~\mu$m and the lowest $A_b$ is from the cloud-free case. }
    \label{fig:albedo_gcm}
    \end{center}
\end{figure}

Thermal phase-curve analysis suggests that a fraction of hot Jupiters exhibit unusually high Bond albedos $A_b$ \citep[e.g.,][]{Schwartz2015}. However, the sources of the reflectivity are unknown, and relatively hot planets are thought to have negligible Bond albedo because the absorption of stellar flux becomes significantly enhanced as the atmospheric temperature increases {\rev \citep[e.g.,][]{parmentier2015}.} We calculated the Bond albedo from our GCMs by diagnosing the top-of-atmosphere reflective flux and the in-coming stellar flux $A_b = \frac{\int_\mathcal{A}\int_{\lambda} F_{\rm ref}(\mathcal{A},\lambda)d\mathcal{A}d\lambda}{\int_\mathcal{A}\int_{\lambda} F_{\rm in}(\mathcal{A},\lambda)d\mathcal{A}d\lambda}$ where $\mathcal{A}$ represents area, and show that clouds do not supply a large Bond albedo. Bond albedos calculated from cloud-free and cloudy ($r_0=5$ and $1~\mu{\rm m}$) GCMs with different $T_{\rm eq}$ are shown in Figure \ref{fig:albedo_gcm}, which shows {\rev the directly diagnosed albedos for each sequence}. At $T_{\rm eq}\ge2000$ K, the Bond albedos are negligible and below 0.01. The low Bond albedo is not surprising, as cloud coverage is limited on the daysides over the temperature range of $T_{\rm eq}=1600$ K to 2800 K, providing a negligible source of reflectivity. 

Recent interpretations of phase-curve observations using idealized energy balance models indicate nontrivial Bond albedos for some hot Jupiters \citep[e.g.,][]{cowan2011,Schwartz2015,Splinter2025}, which are not expected by radiative transfer and GCM calculations \citep[e.g.,][]{parmentier2016}.  Using the energy balance model of \cite{cowan2011}, we re-derive Bond albedos $A_b$ of the GCMs based on the simulated thermal phase curves. By comparing the derived $A_b$ to the ``true" ones obtained from GCM diagnosis, we could assess the potential biases of using energy balance models in interpreting thermal phase curves, and may provide explanations on why (unphysically) high Bond albedos were derived for a subset of hot Jupiters. 
{\rev Following the power-weighted mean method in \cite{Schwartz2015}, the day- and night-side temperatures are obtained from the day- and night-side spectra as 
\begin{equation}
\begin{split}
    & T_d = (\sum_i T_{d,i}w_i)/\sum_i w_i, \\
    & T_n = (\sum_i T_{n,i}w_i)/\sum_i w_i,
\end{split}    
\end{equation}
where $i$ is the index for band passes of different instruments, $T_{d,i}$ is the ``measured" photometric dayside brightness temperature passing through a particular instrumental filter $\phi_i(\lambda)$, and is obtained the same way as those shown in Figure \ref{fig:day-night}, and $w_i$ is the weight for brightness temperatures by the expected integrated power in that band pass $i$, $w_i=\int_{\lambda}B_{\lambda}(T_{est})d\lambda$, $T_{est}=(1/2)^{1/4}T_0$. The Bond albedo can then be inferred using these temperatures and energy balance model \citep{cowan2011,Schwartz2015}. In addition, the heat redistribution coefficient $\epsilon$ (ranges from 0 to 1) can also be defined in their model; however, we do not discuss $\epsilon$ because, unlike the Bond albedo, there is no rigorous comparison between that and the GCM diagnoses. }

\begin{figure*}
    \begin{center}
    \includegraphics[width=1\textwidth]{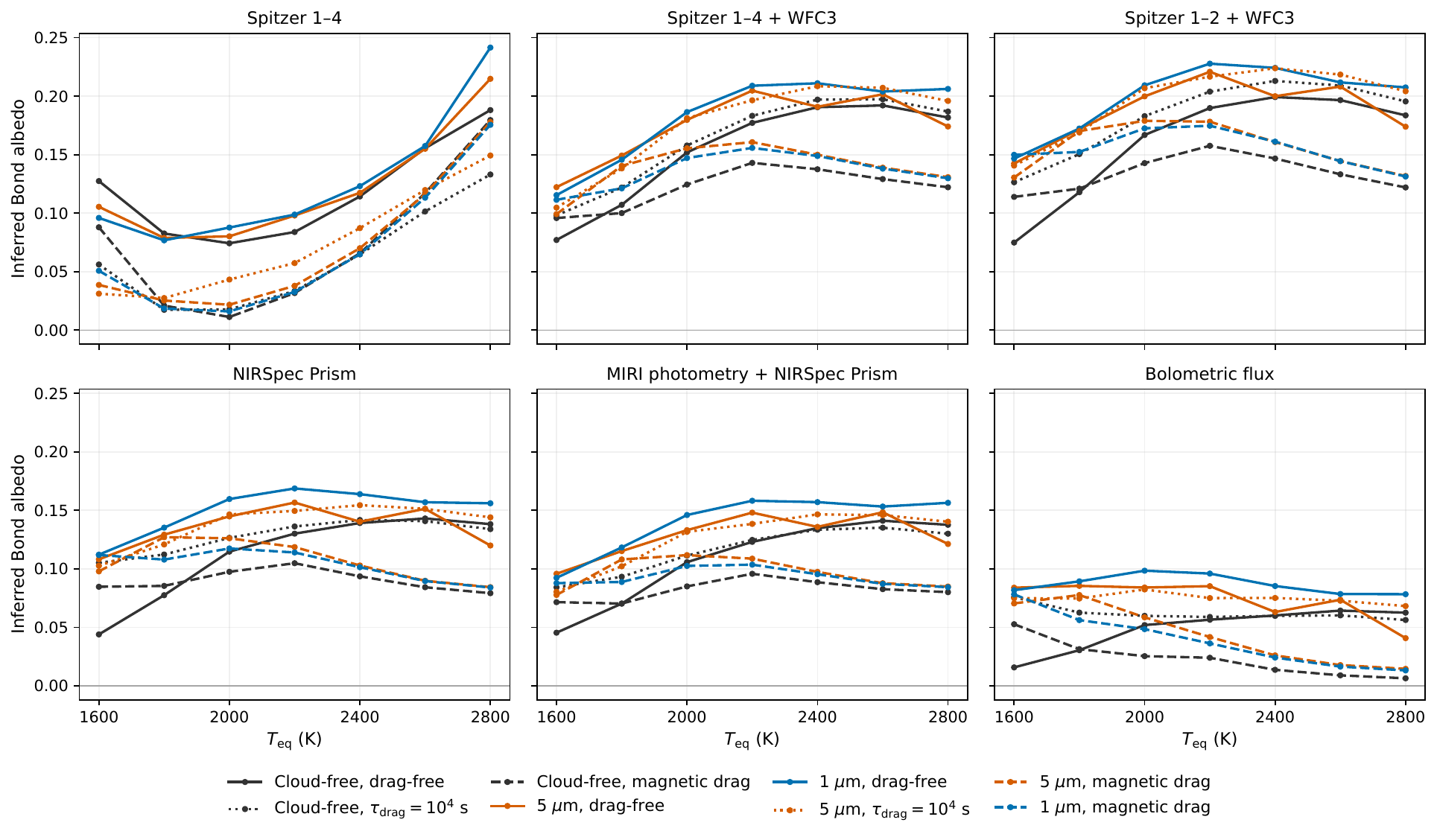}
    \caption{\rev Bond albedos inferred from simulated dayside and nightside thermal emission using the energy-balance model, shown as a function of planetary equilibrium temperature. The first five panels use power-weighted mean brightness temperatures from the indicated combinations of instrumental filters. ``Spitzer 1--4" denotes the four IRAC channels centered at 3.6, 4.5, 5.8, and 8.0 $\mu\mathrm{m}$. ``MIRI photometry" comprises four separate imaging filters: F560W, F770W, F1000W, and F1280W. The final panel uses bolometric dayside and nightside fluxes. Colors distinguish cloud-free models (black) and cloudy models with reference particle radii of \(r_0=5\,\mu\mathrm{m}\) (orange) and \(1\,\mu\mathrm{m}\) (blue); line styles distinguish drag-free models (solid), uniform drag with \(\tau_{\mathrm{drag}}=10^4\) s (dotted), and {\rev magnetic drag with $\tau_{\min}=10^2$~s (dashed)}. These inferred albedos can be compared with the values directly diagnosed from reflected stellar flux in Figure \ref{fig:albedo_gcm} to assess biases associated with limited wavelength coverage and the energy-balance approximation. }
    \label{fig:albedo}
    \end{center}
\end{figure*}

{\rev Figure~\ref{fig:albedo} shows how the inferred Bond albedo depends on wavelength coverage. At $T_{\rm eq}=2800$~K, the four Spitzer channels yield $A_b=0.13$--$0.24$ across the model sequences, compared with $A_b<0.01$ from reflected stellar flux (Figure~\ref{fig:albedo_gcm}). As discussed by \cite{parmentier2021} (their Section~8.1 and Figure~19), brightness temperatures over much of the 3--6~$\mu$m range can lie below the effective temperature in a non-grey atmosphere. Underestimating the outgoing thermal flux in this way leads the energy-balance model to infer a larger reflected fraction.}

{\rev In the final panel we integrate the dayside and nightside thermal spectra over the full modeled wavelength range and use the resulting disk-integrated fluxes in the same energy-balance calculation. At 2800~K the inferred albedos fall to $0.006$--$0.078$. The reduction supports broader wavelength coverage in thermal phase curves, particularly at shorter wavelengths where UHJs emit a substantial fraction of their flux. A considerable residual bias across UHJs persists for some sequences even with bolometric fluxes, illustrating the additional uncertainty in converting two disk-integrated measurements into a global energy budget.}


\section{Conclusions}
\label{sec:con}

This study investigates the influence of clouds on atmospheric circulation and examines whether clouds can account for the observed Spitzer nightside brightness temperatures across a wide range of equilibrium temperatures for hot and ultra-hot Jupiters. To do so, we developed a general circulation model (GCM) that incorporates a self-consistent cloud formation scheme, a realistic radiative transfer module, hydrogen dissociation thermodynamics, and consistent post-processing radiative transfer. Additionally, magnetohydrodynamic dissipation of large-scale flows is parameterized as drag. The model is applied to hot and ultra-hot Jupiters with equilibrium temperatures $T_{\rm eq}$ ranging from 1600 K to 2800 K. Building on our earlier work \citep{tan2024}, this research represents a step forward along the lines of studies such as \cite{perna2012,perezbecker2013,komacek2016,parmentier2016,komacek2017,zhang2017,tan2019uhj,parmentier2021,Baeyens2021,roman2021,Helling2023,roth2024,kennedy2025}, which have used GCM experiments over broad parameter spaces to identify the key physical processes shaping circulation, heat transport, and phase-curve properties in hot Jupiter atmospheres.

We have reached the following key conclusions:
\begin{enumerate}
    \item Clouds form predominantly on the nightside, while daysides remain largely cloud-free across the simulated equilibrium temperature range ($T_{\rm eq}=1600$-2800 K, Figure \ref{fig:maps}). In the absence of atmospheric drag, clouds with characteristic sizes of $r_0 = 1~\mu\text{m}$ and $5~\mu\text{m}$ enhance the amplitude and reduce the eastward offset of thermal phase curves {\rev over much of the modeled spectrum, particularly} where gaseous opacity is low. These effects are strongest at relatively low $T_{\rm eq}$ and gradually weaken as $T_{\rm eq}$ increases (Figure \ref{fig:offset}). However, for ultra-hot Jupiters ($T_{\rm eq} \gtrsim 2200~\text{K}$), the nightside brightness temperatures predicted by drag-free cloudy models remain systematically higher than those observed in the Spitzer 4.5-$\mu$m band. This occurs because strong equatorial jets develop, maintaining a hot equatorial region and clearing clouds near the nightside equator at $T_{\rm eq}\gtrsim 2200$ K, with thick clouds instead forming mainly at mid-to-high latitudes. These patchy nightside clouds cannot sustain low radiative fluxes as temperatures rise. Consequently, incorporating clouds alone does not reproduce the nearly flat trend of nightside brightness temperatures seen in Spitzer 4.5-$\mu$m observations. Instead, cloudy GCMs show a steeper dependence of nightside brightness temperature on $T_{\rm eq}$ than cloud-free models (Figure \ref{fig:day-night}), driven by the increasingly patchy nightside cloud coverage as equilibrium temperature increases.
    \item We discussed two mechanisms responsible for reducing the thermal phase-curve offsets by clouds: the dynamical and geometric effects. The former reduces the eastward shifts of the standing wave structures, {\rev possibly because clouds shorten the radiative timescale at the thermal photosphere} (Figure \ref{fig:ro_div}). The latter, identified in \cite{parmentier2021}, reduces the phase-curve offsets by increasing the brightness contrast between the cloud-free dayside and cloudy nightside. As shown in Figure \ref{fig:offset_nc}, the dynamical effect is at play when cloud particles are either small or large, and the geometric effect is prominent only when the cloud particles are relatively small ($r_0= 1~\mu$m in our models). 
    \item Motivated by the fact that the partial cloud cover on the nightside in drag-free GCMs---likely due to hot equatorial regions---leads to overly high nightside brightness at $T_{\rm eq} \ge 2200$ K, we introduce magnetohydrodynamic dissipation of the large-scale flow, parameterized as drag in our GCM. This drag acts to suppress wave and jet formation, thereby promoting greater cloud coverage on the nightside and further reducing the nightside thermal flux (Figure \ref{fig:maps_drag}). {\rev With a minimum magnetic drag timescale of $10^2$~s, the cloud-free 4.5-$\mu$m nightside brightness temperature stays within $\sim$1190-1200~K over $T_{\rm eq}=1800$-2800~K. Clouds further lower the brightness at fixed irradiation. The $10^4$~s floor yields a steeper brightness-temperature trend but allows upper-atmosphere limb blueshifts of several km~s$^{-1}$ to persist, whereas the default drag strongly damps the net approaching flow in the hotter models (Figures~\ref{fig:T_night} and~\ref{fig:los_velocity}).}
    \item Nightside spectra offer a promising avenue for breaking the degeneracies inherent in broadband photometric measurements. Cloudy GCMs generally exhibit more muted spectral features than their cloud-free counterparts, despite potentially comparable photometric brightness temperatures depending on specific parameters (Figure \ref{fig:spectra2000}). {\rev Wavelength-dependent phase offsets also distinguish models with comparable broadband nightside brightness.  In extreme cloudy cases, e.g. a model with $r_0=1~\mu$m and $T_{\rm eq}=2000$~K, the nightside thermal spectrum can exhibit emission features, most likely through strong thermal scattering (Figure~\ref{fig:cloud_contributions}).}
    \item GCM simulations yield negligible Bond albedos due to limited dayside cloud coverage (Figure \ref{fig:albedo_gcm}).  But interpretations of simulated thermal phase curves using simple energy balance models {\rev in near-IR and IR bands} tend to overestimate the Bond albedos by an order of magnitude or more (Figure \ref{fig:albedo}). {\rev Using bolometric thermal fluxes moves the inferred albedos of UHJs closer to their true values, but systematic overestimations  persist.}
\end{enumerate}

Several key processes in our model rely on parameterizations, which may influence the GCM results. A prominent example is the treatment of cloud particle properties. Moving toward physically motivated cloud formation models with detailed microphysics suitable for GCMs is therefore an important goal. Although portable cloud microphysics schemes have recently been developed \citep{lee2023hj}, their application across a broad range of planetary parameters and validation against observations remain ongoing efforts. {\rev The differing sensitivities of nightside brightness temperatures and limb velocities to the minimum drag timescale also motivate testing magnetic-interaction prescriptions against both thermal phase curves and phase-resolved transmission spectra.} Moreover, disequilibrium chemistry is likely prevalent even in ultra-hot Jupiters and could significantly affect nightside chemistry and thermal emission \citep{tsai2024}. Consequently, incorporating flexible chemical reaction networks (e.g., \citealp{lee2023mini, lee2024bd}) into our GCM framework is a key priority. This development is essential for the accurate interpretation of phase-resolved spectra across the entire planetary orbit.\\

\noindent\textbf{Data Availability.}
The time-averaged general circulation model outputs, phase-resolved
PICASO thermal emission spectra, and Python scripts for reproducing the
figures in this paper are available on Zenodo at
\dataset[https://doi.org/10.5281/zenodo.24358795]{https://doi.org/10.5281/zenodo.24358795}.

\section*{Acknowledgments}
We thank the reviewer for fruitful comments. XT is supported by the National Natural Science Foundation of China (grant No. 42475131) and a Shanghai Jiao Tong University start-up grant.  YL is supported by the National Natural Science Foundation of China (grant No. 42508019). JZ acknowledges support by the 2026 TDLI Winter Intensive Study Program (WISP) in Astrophysics.
This project is supported in part by the Office of Science and Technology, Shanghai Municipal Government (grant Nos. 24DX1400100, ZJ2023-ZD-001). We acknowledge the computational support provided by the Siyuan-1 cluster, supported by the Center for High Performance Computing at Shanghai Jiao Tong University. AI tools were used to assist with coding for generating figures.

\bibliography{draft}{}
\bibliographystyle{aasjournal}



\end{CJK*}
\end{document}